\documentclass[aps,prd,twocolumn,floatfix,nofootinbib,superscriptaddress,tightenlines]{revtex4}
\usepackage[dvipsnames]{xcolor}
\usepackage{ragged2e}
\usepackage{soul}
\usepackage{amsmath}
\usepackage{dcolumn}
\usepackage{lipsum}
\usepackage{amssymb}
\usepackage{url}
\usepackage{epsfig}
\usepackage{graphicx}
\usepackage{amsmath}
\usepackage{bm}
\usepackage{setspace}
\usepackage{lscape}
\usepackage{amsthm}
\usepackage{bbold}
\usepackage{dcolumn}
\usepackage{epsfig}
\usepackage{graphics}
\usepackage{graphicx}
\usepackage{subcaption}
\usepackage[justification=raggedright,singlelinecheck=false]{caption}
\usepackage{etoolbox}
\makeatletter
\patchcmd{\@makecaption}
  {\@tempdima\hfil}
  {\@tempdima\z@}
  {}{}
\makeatother
\usepackage{bm}
\usepackage{xspace}
\usepackage{cancel}
\usepackage{float}
\usepackage{multirow}
\definecolor{darkgreen}{rgb}{0,0.5,0}
\definecolor{purple}{rgb}{0.5,0,0.5}
\definecolor{nblue}{rgb}{0.0,0.0,0.50}
\definecolor{scarlet}{rgb}{1.0,0.2,0}
\definecolor{darkmagenta}{rgb}{0.55, 0.0, 0.55}
\definecolor{darkolivegreen}{rgb}{0.33, 0.42, 0.18}
\definecolor{darkcandyapplered}{rgb}{0.64, 0.0, 0.0}

\usepackage[colorlinks=true, pdfstartview=FitV, linkcolor=purple, citecolor= purple, urlcolor=blue]{hyperref}
\usepackage[normalem]{ulem}

\newcommand{\be}{\begin{equation}}
\newcommand{\tu}{\textcolor{red}{u}}

\newcommand{\fu}{\textcolor{blue}{\bar{f_2}}}
\newcommand{\fd}{\textcolor{blue}{f_1}}

\newcommand{\Me}{\textcolor{blue}{V}}

\newcommand{\Mav}{\textcolor{blue}{{AV}}}
\newcommand{\De}{\textcolor{blue}{DV}}

\newcommand{\Dav}{\textcolor{blue}{{DAV}}}

\newcommand{\Jpsi}{\textcolor{blue}{J/\Psi}}

\newcommand{\td}{\textcolor{darkcandyapplered}{d}}
\newcommand{\tb}{\textcolor{blue}{b}}
\newcommand{\tc}{\textcolor{darkmagenta}{c}}
\newcommand{\ts}{\textcolor{darkgreen}{s}}
\newcommand{\ee}{\end{equation}}
\newcommand{\bea}{\begin{eqnarray}}
\newcommand{\eea}{\end{eqnarray}}
\newcommand{\beas}{\begin{eqnarray*}}
\newcommand{\eeas}{\end{eqnarray*}}
\newcommand{\nn}{\nonumber}

\newcommand{\MeV}{\text{MeV}} 
\newcommand{\GeV}{\text{GeV}} 
\newcommand{\rmh}{\hat{\alpha}_{\mathrm {IR}}}

\newcommand{\eqn}[1]{Eq.~(\ref{#1})}

\begin{document}
\title{Unified Analysis of Screening Masses for Vector and Axial-Vector Mesons and Their Diquark Partners in the Contact Interaction Model}
\author{L. X. Guti\'errez-Guerrero}
\email{lxgutierrez@secihti.mx}
\affiliation{SECIHTI-Mesoamerican Centre for Theoretical Physics,
Universidad Aut\'onoma de Chiapas, Carretera Zapata Km.~4, Real
del Bosque (Ter\'an), Tuxtla Guti\'errez, Chiapas 29040, M\'exico}

\author{M. A. Ramírez-Garrido}
\email{miguelramirez.fcfm@ms.uas.edu.mx}
\affiliation{Facultad de Ciencias F\'isico-Matem\'aticas, Universidad Aut\'onoma de Sinaloa, Ciudad Universitaria, Culiac\'an, Sinaloa 80000,
M\'exico}

\author{M. A. Pérez de León}
\email{marioaldair.perez@uas.edu.mx}
\affiliation{Facultad de Ciencias F\'isico-Matem\'aticas, Universidad Aut\'onoma de Sinaloa, Ciudad Universitaria, Culiac\'an, Sinaloa 80000,
M\'exico}
\affiliation{Facultad de Ciencias de la Tierra y el Espacio, Universidad Aut\'onoma de Sinaloa, Ciudad Universitaria, Culiac\'an, Sinaloa 80000,
M\'exico.}

\author{R. J. Hern\'andez-Pinto}
\email{roger@uas.edu.mx}
\affiliation{Facultad de Ciencias F\'isico-Matem\'aticas, Universidad Aut\'onoma de Sinaloa, Ciudad Universitaria, Culiac\'an, Sinaloa 80000,
M\'exico}


\begin{abstract}
We present a comprehensive study of the screening masses of vector and axial-vector mesons and their corresponding diquark partners within a symmetry-preserving vector–vector contact interaction approach. Our analysis includes mesons and diquarks composed of both light and heavy quarks, providing a unified description of their thermal behavior. The longitudinal and transverse modes of the screening masses are analyzed, and the results are systematically compared with other theoretical approaches. At 
$T=0$ MeV, our predictions agree with available experimental data, and a comparison with the expected free theory limit at high temperatures is also presented. Notably, the parity partners of the lightest mesons and diquarks converge at high temperatures, signaling chiral symmetry restoration within this framework. These results provide a consistent and detailed picture of meson and diquark properties at finite temperature and lay the groundwork for extending the capabilities of the model to baryon screening masses in the quark–diquark picture.
\end{abstract}


\maketitle

\section{Introduction}
 The behavior of hadrons changes drastically at high temperatures, evolving from bound states into a phase characterized by deconfined quark and gluon degrees of freedom. This novel state of matter is known as the quark–gluon plasma (QGP). The study of mesons and baryons, as bound states of quarks, therefore plays a central role in understanding this phenomenon, through both experimental efforts and theoretical analyses.

On the experimental side, efforts to recreate the extreme conditions of the early universe have motivated the construction of advanced accelerator facilities and large-scale programs devoted to relativistic heavy-ion collisions. Among the most notable are the Super Proton Synchrotron  and the Large Hadron Collider at CERN, the Relativistic Heavy Ion Collider at Brookhaven National Laboratory , as well as the more recent Nuclotron-based Ion Collider Facility at the Joint Institute for Nuclear Research  and the Facility for Antiproton and Ion Research at GSI. These experimental efforts are specifically designed to probe the behavior of hadrons under extreme conditions, with particular emphasis on quarkonium states.

From the theoretical perspective, while the thermal evolution of hadronic states in the light-quark sector has been extensively investigated, see Refs. \cite{Chen:2024emt,Mukherjee:2008tr,Dosch:1988vt,Ayala:2012ch,Hatsuda:1992bv,Wang:2015ynf}, studies of the heavy-quark sector remain relatively scarce, despite their critical importance. Quarkonia formation and dissociation provide clear signatures of the QGP, making the precise determination of their screening masses essential for heavy-ion collision phenomenology \cite{Brambilla:2010cs}. Furthermore, quarkonium suppression has long been proposed as an indicator of deconfinement \cite{Matsui:1986dk}. Thus, exploring the behavior of quarkonium under extreme conditions is directly tied to understanding the properties of strongly interacting matter in the early universe.

Mesons, as bound states of two quarks—either light or heavy—play a central role in these investigations.
In fact, it is well established that the first phenomenological order parameter for quark–gluon deconfinement was introduced in the vector (V) channel ($J^{PC} = 1^{--}$) through thermal QCD sum rules, as reported in Ref.~\cite{Bochkarev:1985ex}. This pioneering work was followed by numerous studies \cite{Dosch:1988vt,Ayala:2012ch,Hatsuda:1992bv,Wang:2015ynf}, and a broad spectrum of theoretical models has since been developed to describe V mesons at finite temperature. These approaches include effective chiral Lagrangians \cite{Song:1995ga}, the sigma model \cite{Pisarski:1995xu}, the Kroll–Lee–Zumino model \cite{Hernandez:2025inu}, the Schwinger-Dyson and Bethe–Salpeter equations (BSE) framework \cite{Gao:2020hwo,Fischer:2018sdj}, the nonlocal Polyakov–Nambu–Jona-Lasinio model \cite{Carlomagno:2019yvi}, Gribov quantization \cite{Sumit:2023hjj}, and lattice QCD simulations \cite{Cheng:2010fe,Mukherjee:2008tr,Bazavov:2020teh,Petreczky:2009at}, among others.

The axial-vector (AV) channel has also received considerable attention. The lightest meson in this sector, the $a_1$, has been analyzed through thermal QCD sum rules to study its evolution at finite temperature \cite{Dominguez:2012um,Hatsuda:1992bv,Dey:1990ba,Mamedov:2021dpv}. In addition, the thermal behavior of heavy V states has been examined: the $J/\Psi$ in Ref.\cite{Dominguez:2010ve}, ground-state bottomonia ($\Upsilon$) in Ref.\cite{Dominguez:2013fca}, and V charmonia and bottomonia through bottom-up Anti-de Sitter(AdS)/QCD approaches in Ref.~\cite{MartinContreras:2021bis}. \\

\begin{figure}[h]
\vspace{-7cm}
       \centerline{
\includegraphics[scale=0.6,angle=0]{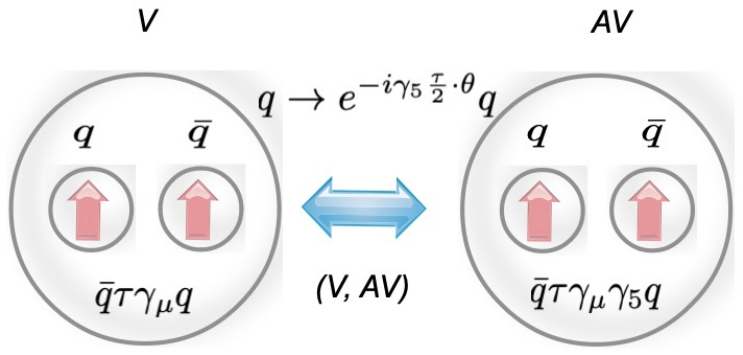}
       }
          \vspace{-6cm}
       \caption{\label{vec-spi-vav} \justifying  Pictorical representation of the relation between chiral partners. Vector and axial-vector mesons are rotated into each other under chiral transformation.} 
\end{figure}

\begin{figure}[h]
   \vspace{-4cm}
   \centering    \includegraphics[scale=0.4,angle=0]{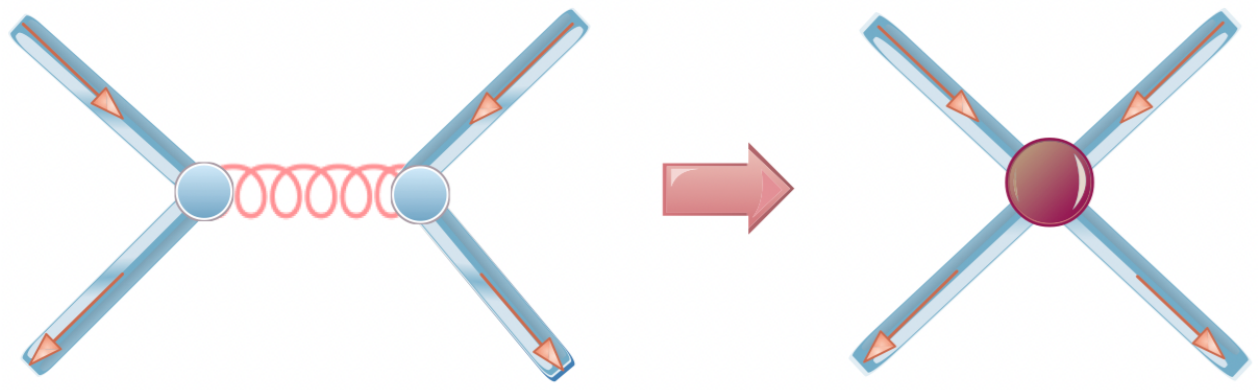}
    \vspace{-4cm}
    \caption{\justifying Diagrammatic illustration of the CI, constructed from the simplified gluon propagator model introduced in Eq.~\eqref{eqn:contact_interaction}.}
    \label{fig:ci}
\end{figure}

An important outcome of QCD sum rule analyses is the contrasting temperature dependence of meson masses: while the $\rho$ meson mass increases with temperature, the $a_1$ mass decreases, leading to a convergence of the two at high temperatures \cite{Dey:1990ba,VeliVeliev:2012cc}. This trend is directly related to the restoration of symmetries in the thermal medium. More generally, it is well established that quarkonia bound states dissolve as the temperature rises, and their masses approach the free-theory limit, $2\sqrt{(\pi T)^2 +M_q^2 }$, independent of the spin–parity structure of the state, with $M_q$ the dressed quark mass \cite{Florkowski:1993bq,Bazavov:2014cta,Karsch:2012na}. In the extreme high-temperature regime and in the limit of massless quarks, this behavior simplifies to the asymptotic form $2\pi T$. A key feature of the V and AV channels is that their lowest-lying states, the $\rho$ and $a_1$ mesons, appear as chiral partners, see Fig. \ref{vec-spi-vav}, reflecting the underlying chiral symmetry of QCD. At zero temperature, there exists a notable mass splitting between these partners, with the $a_1$ meson significantly heavier than the $\rho$, a clear manifestation of spontaneous chiral symmetry breaking in the hadronic spectrum. Understanding how this mass difference evolves with temperature provides crucial insight into the mechanisms of chiral symmetry restoration and the dynamics of hadronic matter under extreme conditions.\\
A natural extension of meson studies is the investigation of diquark correlations, which are central to the description of multiquark bound states, particularly baryons in the quark–diquark picture \cite{GellMann:1962xb,Ida:1966ev,Lichtenberg:1967zz}. Diquarks are color–non-singlet correlations of two quarks and, owing to their color charge, remain confined within hadrons such as baryons, tetraquarks, or pentaquarks, precluding direct experimental observation.
Although diquarks have been widely studied at zero temperature in the context of hadron spectroscopy and nucleon structure \cite{Cahill:1987qr,Cahill:1988dx,Oettel:1998bk,Gutierrez-Guerrero:2019uwa,Gutierrez-Guerrero:2021rsx,Chen:2012qr,Barabanov:2020jvn}, their thermal properties are comparatively less understood. The evolution of diquark correlations with temperature provides insight into the possible persistence or dissolution of baryonic and tetraquark states in the medium, making this analysis a genuine complement to meson studies. Since diquark dynamics directly influence baryon survival and the behavior of hot QCD matter, their role at finite temperature is far from trivial \cite{Chen:2024emt,Wang:2013wk}.
In this work, which also considers diquarks containing heavy quarks, we aim to develop a consistent framework for determining the masses of light and heavy baryons—and possibly tetraquarks \cite{Bedolla:2019zwg}—at finite temperature. Our analysis lays the groundwork for future studies of hadrons under thermal conditions, where the relative contribution of each diquark channel, and its temperature dependence, is expected to play a critical role. Then, this as an initial step toward a systematic and unified understanding of hadronic matter at finite temperature.\\
In this context, we explore the screening masses of mesons and diquarks composed of light and heavy quarks within the contact interaction (CI) framework, where the gluon propagator is assumed to be momentum independent, Fig.~\ref{fig:ci}.
Since its original formulation \cite{Gutierrez-Guerrero:2010waf}, the CI model has demonstrated notable success in reproducing the masses of both baryons and mesons \cite{Gutierrez-Guerrero:2019uwa,Gutierrez-Guerrero:2021rsx,Gutierrez-Guerrero:2024him}, thereby motivating its application to finite-temperature studies. The first such analysis was carried out in Ref.~\cite{Wang:2013wk} and subsequently extended to hadrons with strange-quark content in Ref.~\cite{Chen:2024emt}. Nevertheless, investigations of pseudoscalar and scalar mesons in the heavy-quark sector have remained relatively scarce, as noted in Ref.~\cite{Ramirez-Garrido:2025rsu}. In this work, we build upon the framework established in Ref.~\cite{Ramirez-Garrido:2025rsu} to explore spin-1 mesons, encompassing systems with both light and heavy quarks, thereby offering a natural extension to this class of hadrons.
The article is structured as follows. In Section \ref{CI}, we provide a concise overview of the main features of the CI model and specify the parameters employed in our calculations. Section \ref{Bse-av} focuses on the Bethe–Salpeter equation, with particular attention to spin-1 bound states. In Section \ref{Vector-M}, V mesons are analyzed, while Section \ref{AV-mesons} is devoted to AV mesons. V and AV diquarks are discussed in Section \ref{Diquarks-s}. Finally, Section \ref{Summary} presents a summary of our findings and perspectives for future work.
\section{Contact interaction: key features} \label{CI}
In this section, following the treatment of scalar and pseudoscalar mesons in Ref.~\cite{Ramirez-Garrido:2025rsu}, we outline the key elements of the CI framework relevant to the computation of screening masses for mesons and diquarks. It is important to highlight that at finite temperature, one may in principle distinguish between pole  and screening masses. It should be noted that, away from the limit at $T=0$ MeV, screening masses in a given channel no longer retain a simple or direct correspondence with the pole masses of the bound states present at zero temperature. In the limit $T \to 0$ MeV, the screening mass continuously approaches the pole mass of the ground state, but this connection gradually weakens as thermal effects become dominant. In this work, however, we focus exclusively on screening masses, as they constitute the appropriate quantities extracted from spatial correlation functions and are directly comparable with finite-temperature lattice results.

The remainder of this section introduces the gap equation, specifies the model parameters, describes the temperature dependence of the quark masses, and establishes the notation used throughout our analysis.
\subsection{The Gap Equation at Finite Temperature}
In analogy with the temperature-independent case, the core of our study for $T\neq 0$ MeV lies in the dressed-quark propagator for a quark of flavor $f$, which is determined by solving the gap equation,
\begin{align}
    \nn S_f^{-1}(p;T) &=i \vec{\gamma} \cdot \vec{p} +i \gamma_4 \omega_n +m_f \\
    &+ \frac{16 \pi \hat{\alpha}_{\mathrm{IR}}}{3} \int_{\ell,dq} \gamma_{\mu} S_f(q;T)\gamma_{\mu}\,,
\end{align}
where $p=(\omega_n,\vec{p})$ and $q=(\omega_\ell,\vec{q})$ are the four momenta of the quarks,  $m_f$ the current mass of the quark, and $\omega_n =(2n+1)\,\pi \,T$ is the fermion Matsubara frequency.
Following Ref.~\cite{Chen:2024emt}, we have employed the following notation,
\bea\int_{\ell,dq}:=T\sum^{\infty}_{\ell=-\infty}\int \frac{d^3\vec{q}}{(2\pi)^3}\,.\eea
For the gluon propagator, we have used the CI approach, i.e., 
\begin{eqnarray}
\label{eqn:contact_interaction}
g^{2}D_{\mu \nu}(k)&\to&4\pi\,\hat{\alpha}_{\mathrm{IR}}\,\delta_{\mu \nu}  \,,
\end{eqnarray}
\noindent  where $\hat{\alpha}_{\mathrm{IR}}=\alpha_{\mathrm{IR}}/m_g^2$, with $\alpha_{\mathrm{IR}}$ the running-coupling and \,$m_g=500\,\MeV$  the gluon mass scale~\cite{Aguilar:2017dco,Binosi:2017rwj,Gao:2017uox}. 
The dressed mass $M_f(T)$ is computed through
\begin{align}
\label{dressed mass}
    M_f(T) =m_f +\frac{16 \pi \hat{\alpha}_{\mathrm{IR}}}{3} \int_{\ell,dq} \frac{4M_f(T)}{s_{\ell}+M_f(T)^2}  \, ,
\end{align}
    with $s_\ell=\vec{q}+\omega_\ell$ and, using the Poincaré invariant method, we conclude that,
    \begin{align}\label{gap-T}
        M_f(T) =m_f +M_f(T)\frac{4  \hat{\alpha}_{\mathrm{IR}}}{3\pi} \mathcal{C}^{\rm iu} (M_f(T)^2;T) 
    \end{align}
    where
    \bea
       \nn \mathcal{C}^{\rm iu} (M_f(T)^2;T) = 2 \sqrt{\pi} \, T \, \int_{\tau_{\rm UV}^2}^{\tau_{\rm IR}^2} d\tau \, \frac{e^{- M_f^2 \, \tau} \vartheta_2\left(e^{- 4\pi^2 T^2\, \tau}\right)}{\tau^{3/2}}\,,\\
        \label{newc}
    \eea
with $\vartheta_2(x)$ the Jacobi theta function.
$\tau_{\rm IR, UV}$ are respectively, infrared and ultraviolet regulators or, similarly, $\Lambda_{\rm IR,UV} = \tau_{\rm IR, UV}^{-1}$.  

Before concluding this subsection, let us recall that the present work is restricted to the computation of screening masses. In contrast, determining pole masses at finite temperature requires the reconstruction of a real–time Green function, a substantially more involved procedure. Specifically, one must first compute all relevant screening correlators and form the Fourier series
\begin{equation}
\hat S(\tau) = T \sum_{n=-\infty}^{\infty} e^{-i\nu_n \tau} S(\nu_n),
\end{equation}
where ${\nu_n}$ are the bosonic or fermionic Matsubara frequencies and $S(\nu_n)$ denotes the corresponding thermal Schwinger function. An analytic continuation $\tau \rightarrow \tau + i t$ must then be performed, followed by the limit $\tau \to 0^{+}$, after which suitable step–function–weighted combinations of the resulting expression are required in order to construct the real–time Green function.
Given the complexity of this procedure and its distinct theoretical challenges, the extraction of pole masses lies beyond the scope of the present study.

\subsection{Parameters}
\label{parameters-ci}

For our calculations, we use the parameters from Ref.~\cite{Gutierrez-Guerrero:2021rsx}, where a unified CI-model framework was applied to compute the masses of mesons and baryons with both light and heavy quarks. Following Ref.~\cite{Raya:2017ggu}, and guided by Refs.~\cite{Farias:2005cr,Farias:2006cs}, we introduce a dimensionless coupling $\hat{\alpha}$ defined as a function of
\begin{equation}
\hat{\alpha}(\Lambda_{\mathrm{UV}})=f(\rmh\, , \,\Lambda_{\mathrm{UV}}) , 
\label{eqn:dimensionless_alpha}
\end{equation}
which parametrizes the interaction strength in terms of the ultraviolet cutoff. In close analogy with the running of the QCD coupling with the relevant momentum scale, the dependence of $\hat{\alpha}(\Lambda_{\mathrm{UV}})$ can be described by an inverse logarithmic fit:
\begin{equation}
\label{eqn:logaritmicfit}
\hat{\alpha}(\Lambda_{\mathrm{UV}})=a\,\ln^{-1}\left(\Lambda_{\mathrm{UV}}/\Lambda_0\right),
\end{equation}
where $a = 0.92$ and $\Lambda_0 = 0.36$ GeV~\cite{Raya:2017ggu}. Once a value of $\Lambda_{\mathrm{UV}}$ is specified, this parametrization allows for a direct determination of $\hat{\alpha}(\Lambda_{\mathrm{UV}})$, thereby providing the input needed to construct the parameter set used for the calculation of screening masses within the CI model, as summarized in Table~\ref{parameters}.

 \begin{table}[htbp]
 \caption{ \justifying \label{parameters} 
 Parameters used for the ultraviolet regulator and the coupling constant for different quark combinations  $\hat{\alpha}_{\mathrm {IR}}=\hat{\alpha}_{\mathrm{IRL}}/Z_H$, where $\hat{\alpha}_{\mathrm {IRL}}=4.57$. $\Lambda_{\rm IR}^{(0)} = 0.24$ GeV is a fixed parameter at $T=0$ MeV; for nonzero temperatures this parameter is modified according to Eq.~(\ref{irt}). The value $Z_H = 1$ corresponds to $\alpha_{\rm IR} = 0.93\pi$, which is a model parameter that simulates the zero-momentum strength of a running coupling in QCD~\cite{Chen:2012qr,Gutierrez-Guerrero:2019uwa,Gutierrez-Guerrero:2021rsx,Bedolla:2015mpa}. In contrast, for the heavy sector, Eq.~(\ref{eqn:logaritmicfit}) is used to determine the corresponding parameters.
} 
\begin{center}
\label{parameters1}
\begin{tabular}{@{\extracolsep{0.0 cm}} || l | c | c | c ||}
\hline \hline
 \, quarks \, &\,  $Z_{H}$ \, &\,  $\Lambda_{\mathrm {UV}}\,[\GeV] $ \,  &\,  $\hat{\alpha}_{\mathrm {IR}}$
 \\
 \hline
 \rule{0ex}{2.5ex}
$\, \tu,\td, \ts$ & 1 & 0.905 & 4.57   \\ 
\rule{0ex}{2.5ex}
$\, \tc, \td, \ts $ & \, 3.034 \, & 1.322 & 1.51 \\ 
\rule{0ex}{2.5ex}
$\, \tc $ & \, 13.122 \, & 2.305 & 0.35 \\ 
\rule{0ex}{2.5ex}
$\,  \tb,\tu$, \ts & \, 16.473 \, & 2.522 & 0.28 \\ 
\rule{0ex}{2.5ex}
$\, \tb, \tc$     &  59.056 & 4.131 & 0.08 \\
\rule{0ex}{2.5ex}
$\, \tb $ & 165.848 & 6.559 & 0.03\\
\hline \hline
\end{tabular}
\end{center}
\end{table}

These parameters have also been successfully employed to describe the screening masses of mesons and diquarks in the scalar and pseudoscalar channels, as reported in Ref.~\cite{Ramirez-Garrido:2025rsu}.
It is important to highlight that, in the calculation of hadron masses in their ground and excited states  at $T=0$ MeV, we used a constant $\Lambda_{\rm IR} = 0.24$ GeV, since this infrared regulator implements confinement by
ensuring the absence of quark production thresholds in
all processes. However, for screening masses, this value must vary with temperature \cite{Chen:2024emt,Wang:2013wk}, that is,
$\Lambda_{\rm IR} \to\Lambda_{\rm IR}(T)\,.$
We followed previous works \cite{Chen:2024emt,Wang:2013wk} and defined  the infrarred regulator evolution as,
\bea
\label{irt}
\Lambda_{\rm IR}(T) =\Lambda_{\rm IR}^{(0)} \left(\frac{ M_f(T)}{M_f(0)}\right)^{1/4} \, ,
\eea
 where the value of $\Lambda_{\rm IR}^{(0)}$
  corresponds to the constant infrared scale used at $T=0$ MeV, namely $0.24$ GeV.
With all these parameters in hand, the following subsection explores the behavior of the dressed quark masses for the $\tu$, $\ts$, $\tc$, and $\tb$ flavors\footnote{We include $\td$-quarks by imposing isospin symmetry.} at nonzero temperatures. 
\subsection{Quark Masses}
\label{qmasses}
Table \ref{table-M} establishes the current quark masses used in our calculations at $T=0$ MeV; however, we are now able to compute the dressed quark masses as a function of the temperature using the parameters presented above.
Table~\ref{table-M} presents the results of our calculation using \eqn{gap-T} for the dressed  mass at zero temperature. \\
\begin{table}[h!]
\caption{\justifying \label{table-M}
Current ($m_{f}$) and dressed masses
($M_{f}$) for quarks in GeV obtained using $T=0$ MeV.}
\begin{center}
\begin{tabular}{@{\extracolsep{0.0 cm}} || c | c | c | c || }
\hline 
\hline
 $m_{\tu}=0.007$ &$m_{\ts}=0.17$ & $m_{\tc}=1.08$ & $m_{\tb}=3.92$   \\
 \rule{0ex}{2.5ex}
 $ M_{\tu}=0.367$ \, & \, $  M_{\ts}=0.53$\; \, &\,   $  M_{\tc}=1.52$ \, &\,  $  M_{\tb}=4.75$   \\
 \hline
 \hline
\end{tabular}
\end{center}
\end{table}

\begin{figure}[b]
    \centering    \includegraphics[width=1\linewidth]{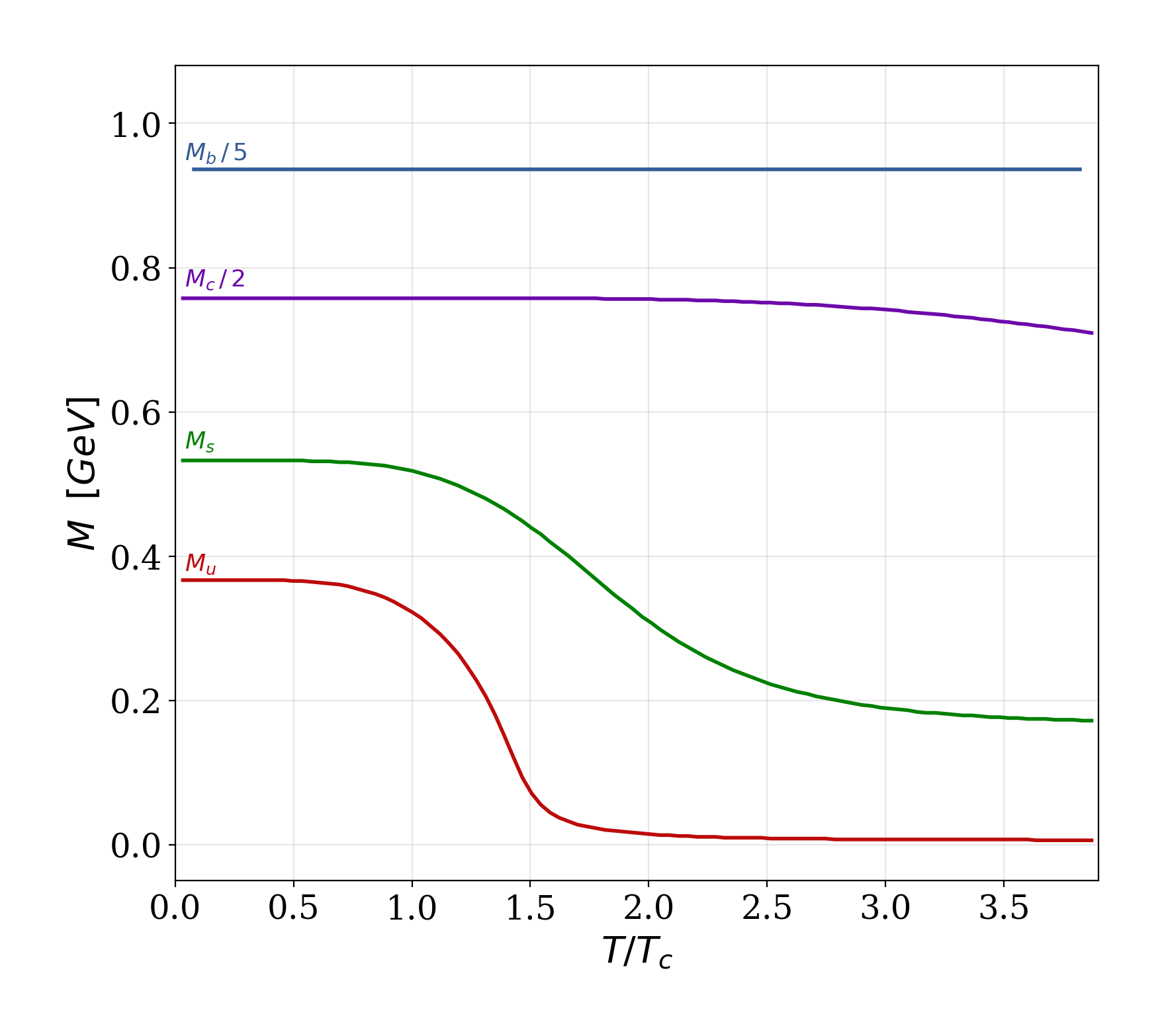}
    \vspace{-1cm}
    \caption{\justifying Results for the screening quark masses computed across a temperature range from $0$ to $4\, T_c$, with $T_c=155$ MeV \cite{Andronic:2017pug,Bazavov:2011nk}. At high temperatures, the quark mass asymptotically tends toward its bare mass, $M(T)\to m$.}
    \label{fig:dressed}
\end{figure}

We illustrate the evolution of the screening masses of these quarks as a function of temperature in Fig.~\ref{fig:dressed}.
In the present work, which includes mesons containing heavy quarks, we adopt $T_c = 155$ MeV, consistent with values used in lattice QCD simulations \cite{Andronic:2017pug, Bazavov:2011nk}, to facilitate comparison with other approaches. However, it is important to emphasize that this temperature acts merely as a rescaling factor in our approach.

It is worth noting that, as the temperature increases, the quark mass decreases—a trend especially pronounced for light quarks, as discussed in Ref. \cite{Ramirez-Garrido:2025rsu}, which also compares the quark screening masses obtained within the CI model with those from other approaches.

\section{Bethe Salpeter Equation}
\label{Bse-av}
\begin{figure}[h!]
   \centering
\includegraphics[scale=0.5,angle=0]{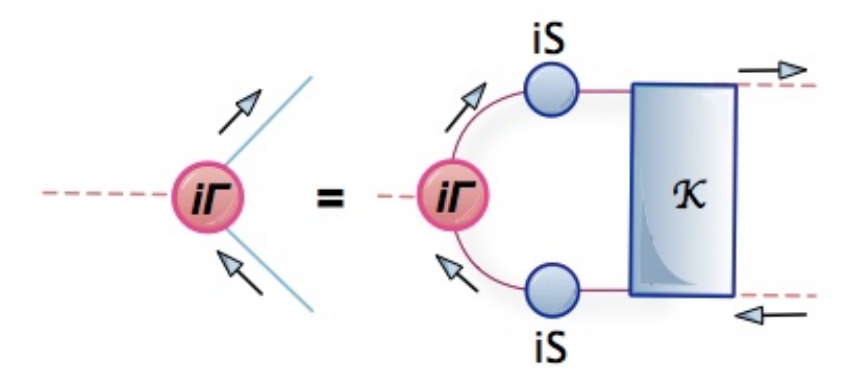}
    \caption{\justifying Diagrammatic representation of the BSE. Blue (solid) circles represent dressed quark propagators $S$, red (solid) circle is the meson BSA $\Gamma$ while the blue (solid) rectangle is the dressed-quark-antiquark scattering kernel ${\mathcal {K}}$.}
    \label{fig:BSEfig}
\end{figure}

Mesons are classified according to their spin and parity into scalar (S), pseudoscalar (PS), V, and AV states. While the screening masses of S and PS mesons have been extensively investigated in both the light- and heavy-quark sectors within the contact interaction model \cite{Ramirez-Garrido:2025rsu,Chen:2024emt}, the present study aims to extend the understanding of V and AV meson screening masses in this framework.
Hadrons formed by two quarks, including mesons and diquarks, can be described as relativistic bound states using the homogeneous BSE, as illustrated diagrammatically in Fig.~\ref{fig:BSEfig}. \\
 
 This equation, in the case of nonzero temperatures, can be mathematically expressed as~\cite{Salpeter:1951sz},
 \begin{equation}
[\Gamma(Q;T)]_{tu} = \int \! \frac{d^4q}{(2\pi)^4} [\chi(Q;T)]_{sr} {\mathcal K}_{tu}^{rs}\,,
\label{genbse}
\end{equation}
where $[\Gamma(Q;T)]_{tu}$ represents the bound-state's Bethe-Salpeter Amplitude (BSA) and $\chi(Q;T) = S(q+Q;T)\Gamma(Q;T) S(q;T)$ is the Bethe Salpeter wave-function; $r,s,t,u$ represent colour, flavor and spinor indices; and ${\mathcal K}$ is the relevant quark-antiquark scattering kernel. In this work, we employ the rainbow--ladder (RL) truncation, which corresponds to the leading-order term in the widely used truncation scheme that preserves global symmetries~\cite{Bender:1996bb,Munczek:1994zz}. Within the RL framework, the quark--antiquark (or quark) two-body scattering kernel takes the simplest form
\begin{equation}
{\mathcal K}_{tu}^{rs}
= g^{2} D_{\mu\nu}(k)\,[i\gamma_{\mu}]_{tu}\,[i\gamma_{\nu}]_{rs}.
\label{eq:kernel}
\end{equation}
Furthermore, the four-momenta $Q^{\mu}=(\vec{Q},\Omega_m)$ contains the temperature dependence through $\Omega_m=2m\pi T$, which is the Boson Matsubara frequency.
Throughout this article, our analysis is restricted to the zeroth Matsubara frequency.
A general decomposition of the BSA for mesons ($\fd\fu$) in the CI is given by,
\begin{align}
    \Gamma_i(Q;T)= \sum_j E_i^{(j)}(T) \, A^{(j)}_i  \,,
\end{align}
with $i$ denoting the type of hadron under study, specifically V and AV mesons and diquarks, and $j=1,\dots,n$ running over the number of Lorentz structures accordingly. In the following, we shall discuss the explicit analytical expressions for each case analyzed in this work. These channels play a central role in the investigation of chiral symmetry restoration, as will be further elaborated below.
\subsection{Vector Mesons}
\label{Vector-M}


In our analysis we consider spin-1 mesons, starting with V states built from heavy ($Q\bar{Q}$), heavy–light ($Q\bar{q}$), and light ($q\bar{q}$) configurations, focusing on the $\tu$, $\td$, $\ts$, $\tc$, and $\tb$ flavors within the $SU(5)$ multiplets of the quark model~\cite{GellMann:1964nj,Zweig:1964jf,Zweig:1964ruk}.\\
In the CI model, which is independent of the relative momentum, only two of the eight Lorentz structures of the V mesons BSA remain nonvanishing~\cite{Llewellyn-Smith:1969bcu,Maris:1999nt}.\\
At finite temperature, the breaking of $O(4)$ symmetry leads to a separation into longitudinal and transverse components in the Bethe–Salpeter equation~\cite{Blaschke:2000gd,Chen:2024emt}.
Consequently, for mesons composed of quarks of flavor $\fd$ and antiquarks of flavor $\fu$ at nonzero temperature, the amplitude can be expressed as
\begin{equation}
\label{bsamvc}
\Gamma_{\Me}(Q^2;T) =
\gamma_4 \, E_{\Me}^{L}(T) +
\gamma_\mu^{\perp}E_{\Me}^{T}(T) \,,
\end{equation}
where $\gamma_\mu^{\perp}(Q) = \gamma_\mu -(\gamma\cdot Q) \, Q_\mu/Q^2$. The term proportional to $\gamma_\mu^\perp$ is related to the transverse mode while results related with $\gamma_4$ are known as longitudinal contributions.
At zero temperature, the equations for the longitudinal and transverse components are naturally identical.
It is straightforward to note that there are now two mass components. The masses associated with the longitudinal ($m_{\Me}^L$) and transverse ($m_{\Me}^T$) components  are determined by the solutions of
\bea
\label{bsevc}
 1-{\mathcal K}_{\Me}^{(L,T)}\left(Q^2=-\left(m_{\Me}^{(L,T)}\right)^2;T\right) &=&0\,
\label{sols}\,,
\eea
where longitudinal and transverse kernels are
\bea
\label{bsekervc}
\nonumber
{\mathcal K}_{\Me}^L(Q^2;T)&= &\frac{2\hat{\alpha}_{\rm IR}}{3\pi}\int^1_0 d\alpha \, ( {\cal L}_{\Me}+2{\mathcal R}^{{\rm{ iu}},L}) \,,\\
{\cal K}_{\Me}^T(Q^2;T) &=&
 \frac{2\rmh}{3\pi}
\int_0^1 d\alpha \, \,\,{\cal L}_{\Me}\,,
\eea
with
\bea  
\nn{\cal L}_{\Me} \equiv {\cal L}_{\Me}(Q^2;T)&=&[M_{\fd} M_{\fu}-\alpha M_{\fd}^2 - (1-\alpha)M_{\fu}^2 \\
\nn &-&2 \alpha (1-\alpha)Q^2]\overline{\mathcal{C}}_1^\mathcal{T}\,, 
\eea
and the function $\overline{\mathcal{C}}_1^\mathcal{T}$ is defined as
\begin{eqnarray}
  \nn \overline{\mathcal{C}}_1^\mathcal{T}&\equiv& \overline{\mathcal{C}}_{1}^{\rm iu}(\omega^{(1)}(T);T)\,,
\end{eqnarray}
with
\begin{eqnarray} 
 \nn   \omega^{(1)}(T)&=&M_{\fu}^2(T)(1-\alpha)+\alpha M_{\fd}^2(T) +\alpha(1-\alpha)Q^2,\\
    \nn
    \overline{\mathcal{C}}_{1}^{\rm iu}(z;T)&=&-(d/dz)\mathcal{C}^{\rm iu}(z;T)\, .
\end{eqnarray}
Additionally, the ${\mathcal R}^{{\rm iu,} L}$ function is defined as
\begin{align}
\nonumber
&{\mathcal R}^{{\rm iu,} L} \equiv {\mathcal R}^{{\rm iu,} L}(\omega^{(1)}(T);T)\\
 = &\int^{1/\Lambda_{\rm IR}^2}_{1/\Lambda_{\rm UV}^2}d\tau {\rm e}^{-\tau \omega^{(1)}(T)}\sqrt{\frac{\pi}{\tau}}\bigg[-\frac{d}{d\tau}-\frac{1}{2\tau}\bigg]2T\vartheta_2({\rm e}^{-\tau4\pi^2T^2})\,.
\end{align}
In the limit when $T\to 0$ MeV, the function ${\mathcal R}^{{\rm iu,} L}$ vanishes, and the transverse and longitudinal components become identical. This follows directly from Eq.~(\ref{sols}), which show that their difference arises solely from this extra term. 
Furthermore, the canonical normalization conditions for the longitudinal and transverse components are 
\bea
\frac{1}{\left(E_{\Me}^{(L,T)}\right)^2} =\frac{9m_G^2}{4\pi\alpha_{\rm IR}}\frac{d}{dQ^2} \mathcal{K}_{\Me}^{(L,T)}(Q^2)\bigg|_{Q^2=-\left(m_{\Me}^{(L,T)}\right)^2},\nn\\
\label{canovc}
\eea
where we have omitted the temperature dependence in the previous expression.
Therefore, by solving Eqs.~(\ref{sols}) and (\ref{canovc}) with the parameters defined in Secs. \ref{parameters-ci} and \ref{qmasses} for the longitudinal and transverse components at $Q^2=-m_V^2$, the screening masses and the BSAs can be extracted.
We present our results for the masses and BSAs at 
$T=0$ MeV in Table \ref{table-mesones-vec}.
\begin{table}[htb]
\caption{\justifying \label{table-mesones-vec}
Computed and experimental masses for V mesons (GeV) and BSA in $T=0$ MeV. Column five shows the reduced mass for each meson, and column six shows the limit
$2\sqrt{(\pi T)^2 +(2 M_R)^2 }$ at $T=500$ MeV.}
\begin{center}
\begin{tabular}{@{\extracolsep{0.3 cm}}ccccc|c}
\hline
\hline
Meson   &Exp.& CI & $E_{\Me}$ & $2M_R $ \,  & Limit\, \,\\ \hline 
\rule{0ex}{2.5ex}
$\rho(\tu\bar{\td})$ & 0.78 & 0.92 & 1.53 & 0.367 \, & 3.23\, \, \\
\rule{0ex}{2.5ex}
$K_1(\tu\bar{\ts})$  & 0.89 & 1.03 & 1.63 & 0.433 \, & 3.26 \, \, \\
\rule{0ex}{2.5ex}
$\phi(\ts\bar{\ts})$ & 1.02 & 1.12 & 1.73 & 0.530 \, &  3.32 \, \, \\
\rule{0ex}{2.5ex}
$D^{*0}(\tc\bar{\tu})$ & 2.01 &  2.05 &  1.23 &  0.591 \, &  3.36\, \, \\
\rule{0ex}{2.5ex}
$D_{\ts}^*(\tc\bar{\ts})$ & 2.11 &  2.13 &  1.31 &  0.785 \,  &  3.51 \, \, \\
\rule{0ex}{2.5ex}
$B^{+*}(\tu\bar{\tb})$ & 5.33 & 5.32 & 0.65 & 0.681 \, & 3.42 \, \, \\
\rule{0ex}{2.5ex}
$B_{\ts}^{0*}(\ts\bar{\tb})$ & 5.42 & 5.41 & 0.66 & 0.953  \, &  3.67 \, \, \\
\rule{0ex}{2.5ex}
$B_{\tc}^*(\tc\bar{\tb})$ & $\cdots$  & 6.30 & 0.27 &  2.300 \, & 5.57 \, \, \\
\rule{0ex}{2.5ex}
$\Jpsi(\tc\bar{\tc})$ & 3.10 & 3.12 & 0.61 & 1.520 \, & 4.37 \, \, \\
\rule{0ex}{2.5ex}
$\Upsilon(\tb\bar{\tb})$ & 9.46 & 9.41 & 0.15 & 4.750 \,  & 10.0 \, \, \, \\
\hline
\hline
\end{tabular}
\end{center}
\end{table}

It is evident that at zero temperature, our results can be compared with experimental data \cite{ParticleDataGroup:2024cfk}, and they show good agreement.
At zero temperature, $E_{\Me}^L = E_{\Me}^T$ and $m_{\Me}^L = m_{\Me}^T$, so in this case the longitudinal and transverse labels are omitted.
For our calculations involving mesons composed of different quark flavors, we modify \eqn{irt} by replacing $M_f$ with the reduced mass $M_R = M_{\fd} M_{\fu}/[M_{\fd} + M_{\fu}]$.
In the last column of the table we report the screening-mass limit at 500 MeV. In this limit, the masses of the light mesons display a percentage increase exceeding 100\% with respect to their $T=0$ MeV values, whereas for the heavy mesons the increase is much smaller, about 6\%.\\

\begin{figure*}[t!]
\begin{tabular}{@{\extracolsep{-2.3 cm}}c}
\renewcommand{\arraystretch}{-1.6} %
 \hspace{-1cm}
\includegraphics[scale=0.65]{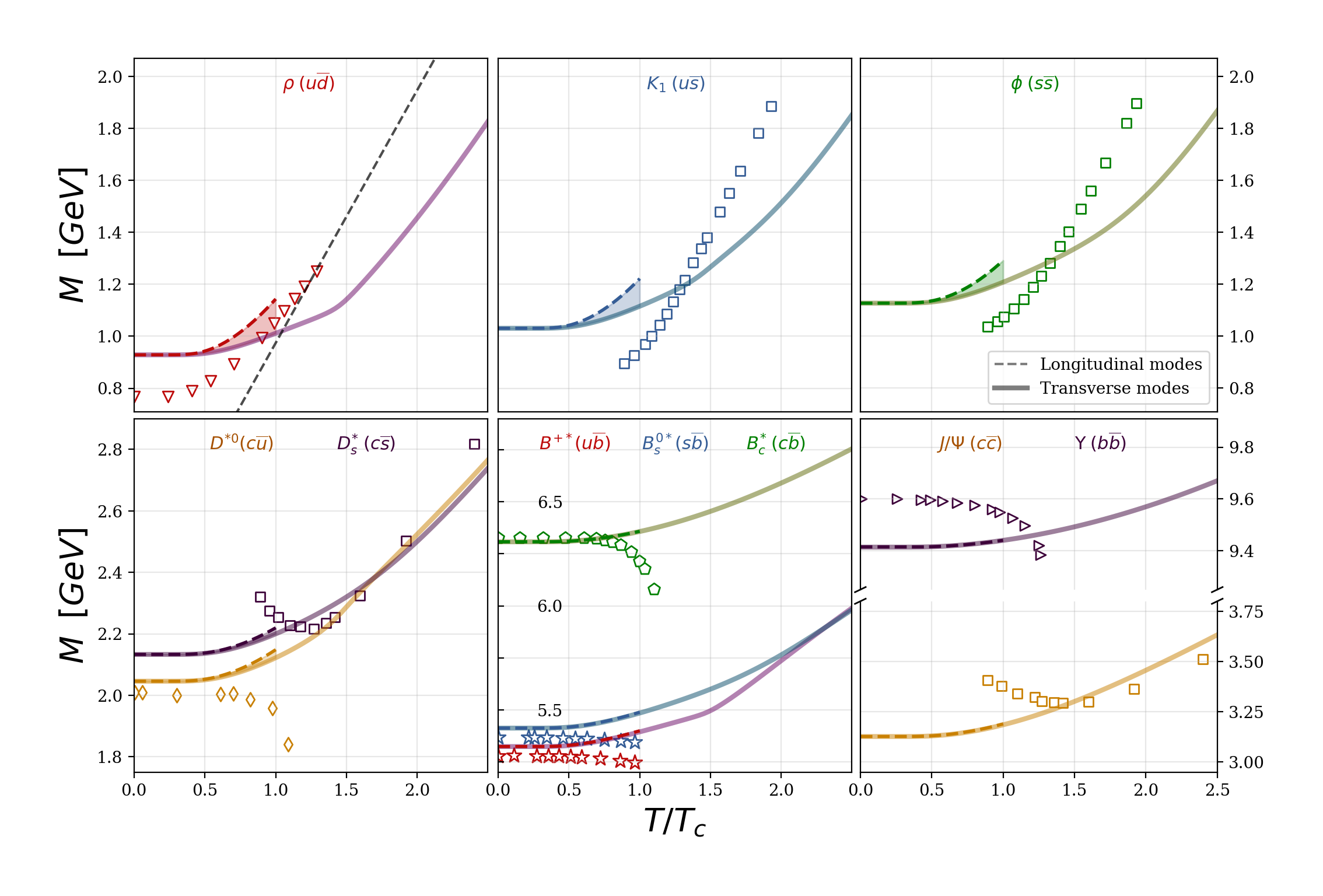}
\end{tabular}
\vspace{-1cm}
\caption{\justifying \label{fig:vectorSM}Longitudinal and transverse screening masses for V mesons composed of the five quark flavors($\tu$, $\td$, $\ts$, $\tc$ and $\tb$). The longitudinal components are shown with dashed lines, while the transverse components are shown with solid lines. 
In the top left panel, the lightest vector meson is shown, the $\rho$ meson, The  black dashed line corresponds to the free-theory limit,  $m = 2\pi T/T_c$; the top center panel displays the $K_1$ screening mass results; the top right panel displays strangeonium vector mesons. The bottom left panel is dedicated to vector $D$-mesons ($D^{*0}$ and $D^{*}_s$); the bottom center panel is committed for vector $B$-mesons ($B^{+*}$, $B^{0*}_s$ and $B^{*}_c$) and, the bottom right panel shows the results for charmonium and bottomonium vector mesons. Results shown with squares correspond to lattice-QCD screening-mass calculations from Ref.~\cite{Bazavov:2020teh}, while stars denote results obtained using an effective chiral Lagrangian approach at finite temperature from Ref.~\cite{Montana:2021vks}. In channels where screening-mass results are not available, we compare our results with finite-temperature meson masses reported in the literature. In particular, lattice-QCD determinations and continuum-model calculations from Refs.~\cite{Dominguez:2013fca,Aarts:2022krz,Song:1993af,Veliev:2011zz} are shown using pentagons, diamonds, inverted triangles, and right-pointing triangles. These latter results do not correspond to screening masses but are included as qualitative benchmarks, providing complementary insight into the thermal behavior of vector mesons in sectors where direct screening-mass calculations are presently unavailable.
}
\end{figure*}

Fig. \ref{fig:vectorSM} shows the evolution of the mass with temperature of V mesons. 
It should be noted that, in our formulation, we do not incorporate the temperature-induced separation between the transverse and longitudinal modes of the gluon propagator. This simplification, widely used in continuum analyses of screening masses, has only a mild impact within the temperature range explored in this work, $T \lesssim 2\, T_c$, as discussed in Ref.~\cite{Cucchieri:2007ta}. For clarity, we also note that the screening-mass curves in Fig.\ref{fig:vectorSM}  are shown up to $2.5\,T_c$ solely to illustrate the qualitative trend predicted by our approximation.
Within the CI framework, both the longitudinal and transverse components increase, but they gradually separate as the temperature rises. This effect is more pronounced for light mesons and less so for heavy mesons. For $\rho$, $K_1$, and $\phi$, the differences between longitudinal and transverse modes exceed 5\% at $T/T_c=0.74$, 0.82, and $0.91$, while they remain below 1.5\% for charmed mesons and under 0.1\% for bottom mesons.
For the lightest mesons, a black dashed line indicates the free-theory limit, which corresponds to the case where the constituent quark masses vanish, providing a reference for the behavior of the screening masses in the chiral limit.
Another noteworthy feature is that the transverse screening masses of V mesons at low temperatures (below $0.4\,T_c$) closely approach a constant value $M(0)$. However, beyond a characteristic temperature $T_H$, the screening mass $M(T)$ increases rapidly, separating from $M(0)$. In particular, $T_H$ satisfies
\begin{equation}\label{th}
    M(0)\left(\frac{dM(T)}{dT}\right)\bigg|_{T=T_H}=0.5 \text{ GeV}.
\end{equation}
The temperatures $T_H$ at which this change occurs for V mesons are listed in Table~\ref{temps vector}.
 \begin{table}[b]
 \caption{ \justifying \label{temps vector} 
The temperature $T_H$ at which the monotonic increase of the screening mass ends and a faster growth sets in, determined using Eq.~(\ref{th}).}
\begin{center}
\label{temp1}
\begin{tabular}{@{\extracolsep{0.0 cm}}  c | c | c | c }
\hline \hline
 \, Meson \, &\,  $T_{H}/T_c$ \, & \, Meson \, &\,  $T_{H}/T_c$ \, 
 \\
 \hline
 \rule{0ex}{2.5ex}
$\, \rho(\tu\overline{\td})$ & 0.53   & $\, B^{+*}(\tu\overline{\tb}) $ & \, 0.37   \,\\ 
\rule{0ex}{2.5ex}
$\,  K_1(\tu\overline{\ts}) $ & \, 0.53 \, & $\, B_{\ts}^{0+}(\ts\overline{\tb}) $ & 0.38 \\ 
\rule{0ex}{2.5ex}
$\,  \phi(\ts\overline{\ts}) $ & \, 0.53  \, & $\, B_{\tc}^*(\tc\overline{\tb}) $ &  0.40 \\ 
\rule{0ex}{2.5ex}
$\,   D^{*0}(\tc\overline{\tu})$ & \, 0.46 \, & $\, \Jpsi(\tc\overline{\tc}) $ & 0.45  \\ 
\rule{0ex}{2.5ex}
$\,  D_{\ts}^{*}(\tc\overline{\ts})$ & 0.47 & $\, \Upsilon(\tb\overline{\tb}) $ &  0.42  \\
\hline \hline
\end{tabular}
\end{center}
\end{table}
Our results reveal the following hierarchy among the 
$T_H$  temperatures,
\bea
\nn &&T_H^{\rho} \approx T_H^{K_1} \approx T_H^{\phi}\,,\\
\nn &&T_H^{D^{*0}} <T_H^{D_{s}^{*}}\,,\\
\nn &&T_H^{B^{+*}} <T_H^{B_{\ts}^{0+}}<T_H^{B_{\tc}^*}<T_H^{J/\Psi}\,.
\eea
From our analysis, we can infer that the change in the behavior of the screening masses occurs more rapidly in mesons composed of the lightest quarks.\\
In addition, we present in Fig.~\ref{fig:vectorSM} a comparison of our results for the transverse component with lattice data from Refs.~\cite{Bazavov:2020teh,Aarts:2022krz}, as well as with predictions obtained from an effective chiral Lagrangian model~\cite{Montana:2021vks,Song:1993af} and from QCD sum rules~\cite{Veliev:2011zz,Dominguez:2013fca}. In most models the mass increases with temperature, but in mesons containing one or two bottom quarks it may decrease, as predicted by QCD sum rules Ref.~\cite{Veliev:2011zz,Dominguez:2013fca} and effective chiral Lagrangian model~\cite{Song:1993af}.
The behavior of the masses in these approaches is similar to our CI results at low
temperatures; however, clear deviations appear once the system reaches and exceeds the critical
temperature. This mainly originates from a limitation of finite-temperature QCD sum rules: correlators are
evaluated only at leading order in perturbative QCD (one loop), and higher-order thermal corrections are not reliable.
Therefore, these calculations are valid only for $0 \leq T \leq T_c$, directly influencing the temperature
dependence of the screening masses, as discussed in Ref.~\cite{Dominguez:2013fca} and confirmed in
Ref.~\cite{Veliev:2011zz}, where it is explicitly stated that the results should not be extended to $T > T_c$.
A similar decreasing trend is observed in the points marked with stars \cite{Song:1993af}. In that work, the
results are only reliable near $T_c$, and for higher temperatures the predictions become ambiguous, since the
model cannot describe the symmetry-restored regime.
This variation among theoretical predictions highlights the fragmented state of the existing literature, particularly in the heavy-meson sector. Since the available results are derived from fundamentally different methodologies and assumptions, their temperature dependences cannot be compared on equal footing. Although we have compiled the broadest set of screening-mass results accessible, they necessarily originate from diverse approximations, which accounts for the discrepancies observed. The present work seeks to address this issue by offering a unified analysis within a single, consistent theoretical framework.\\
In our framework, the longitudinal and transverse channels indeed decouple at finite temperature; however, their numerical behavior differs significantly. In particular, as already observed in Dyson--Schwinger equation studies truncated at the rainbow--ladder level \cite{Maris:2001rq,Blaschke:2000gd}, the longitudinal mode becomes unstable with respect to $q\bar q$ dissociation already around $T \sim 180\,\text{MeV}$. The same behavior is found in our contact interaction model, although the longitudinal equation remains formally well defined above $T_c$, the corresponding solution no longer represents a stable correlation and cannot be reliably extracted.\\
Also in several lattice QCD studies, only the transverse component of the screening mass is reported. This choice is motivated by the fact that, at high temperature or nonzero chemical potential, the longitudinal propagator is strongly suppressed in the infrared, whereas the transverse one remains comparatively stable \cite{Bazavov:2019www}. Moreover, the transverse channel is usually determined with better numerical precision in lattice simulations, making it the most reliable observable.
Thus, the results presented in this work provide the CI model prediction for the transverse component over a broader temperature range than for the longitudinal one, which is calculated only up to $T_c$. This extension constitutes a valuable step toward a more complete description, as it allows for analytical expressions of screening masses and BSAs. However, the same analytical treatment also increases the complexity of incorporating longitudinal effects beyond $T_c$, making predictions above this value less robust. Nevertheless, the present results offer a consistent baseline that can guide future refinements and extensions of the approach.

In the following section, we focus on spin-1 mesons with positive parity, namely the AV mesons, and analyze their properties in detail within our framework.

\subsection{Axial-Vector Mesons}
\label{AV-mesons}
\begin{figure*}[ht]
\begin{tabular}{@{\extracolsep{-2.3 cm}}c}
\renewcommand{\arraystretch}{-1.6} %
 \hspace{-1cm} \includegraphics[scale=0.65]{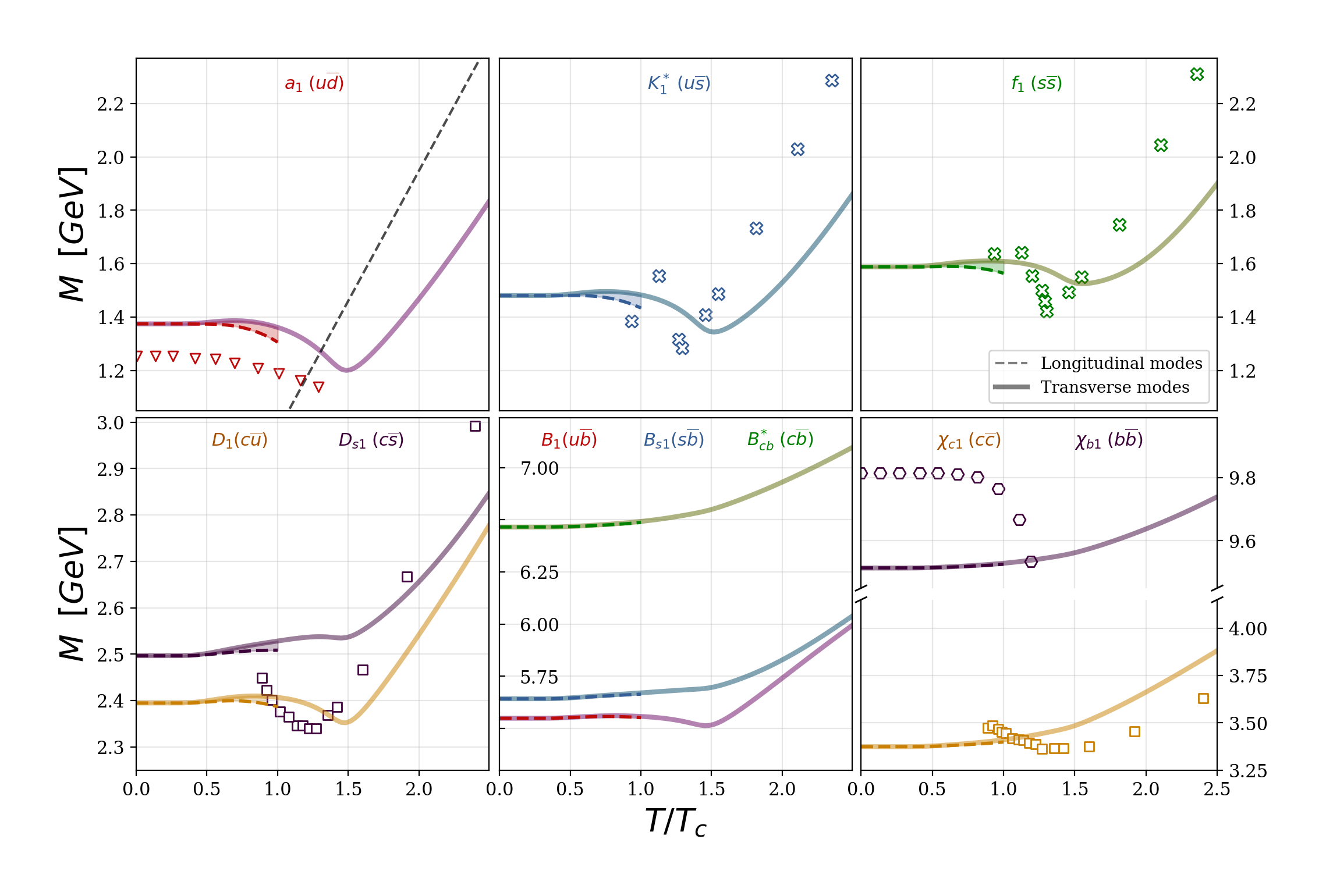}
\end{tabular}
\vspace{-1cm}
\caption{ \justifying \label{fig:av}Longitudinal and transverse screening masses for AV mesons composed of the five quark flavors($\tu$, $\td$, $\ts$, $\tc$ and $\tb$).The longitudinal components are shown with dashed lines, while the transverse components are shown with solid lines. 
In the top left panel, the lightest axial-vector meson is shown, the $a_1$ meson, The black dashed  line corresponds to the free-theory limit,  $m = 2\pi T/T_c$; the top center panel displays the $K_1^*$ screening mass results; the top right panel displays strangeonium axial-vector mesons. The bottom left panel is dedicated to axial-vector $D$-mesons ($D_1$ and $D_{\ts 1}$); the bottom center panel is committed for axial-vector $B$-mesons ($B_1$, $B_{\ts 1}$ and $B^{*}_{\tc\tb}$) and, the bottom right panel shows the results for charmonium and bottomonium axial-vector mesons. Squares and crosses denote lattice results from Refs.~\cite{Bazavov:2020teh,Cheng:2010fe}, inverted triangles represent results from an effective chiral Lagrangian in Ref.~\cite{Song:1993af}. Hexagons indicate QCD sum-rule predictions based on finite-temperature meson masses for the
heaviest mesons from Ref.~\cite{Yazici:2016foi}. Although these results do not correspond to screening masses, they
provide a useful benchmark in sectors where screening-mass calculations are not yet available.
}
\end{figure*}

In analogy to the V case, the transverse and longitudinal components are present, and the BSA for an AV meson composed of quarks with flavour $\fd$ and antiquarks with flavor $\fu$ at non-zero temperature is given by
\begin{equation}
\label{bsamvc}
\Gamma_{\Mav}(Q^2;T) =
\gamma_5\gamma_4 \, E_{\Mav}^{L}(T) +
\gamma_5\gamma_\mu^{\perp}E_{\Mav}^{T}(T)
\end{equation}
and the corresponding eigenvalue equation read as,
\bea\label{eigenAV}
1+\mathcal{K}_{\Mav}^{(L,T)}\left(Q^2=-\left(m_{\Mav}^{(L,T)}\right)^2;T\right) =0 
\eea
where the $\Mav$-kernel takes the form
\bea
\nn    \mathcal{K}_{\Mav}^L(Q^2;T)&=& \frac{2\rmh}{3\pi}
\int_0^1 \,d\alpha\,\,
\left( \mathcal{C}_1^\mathcal{T} +
{\cal L}_{\Mav} \, 
\overline{\mathcal{C}}_1^\mathcal{T} +2{\mathcal R}^{\rm iu,L}\right)\, , \\
 \mathcal{K}_{\Mav}^T(Q^2;T)&=& \frac{2\rmh}{3\pi}
\int_0^1 \,d\alpha\,\,
\left( \mathcal{C}_1^\mathcal{T} +
{\cal L}_{\Mav} \, 
\overline{\mathcal{C}}_1^\mathcal{T} \right) \, ,
\eea
with,
\bea  
\hspace{-4mm}\nn {\cal L}_{\Mav}\equiv {\cal L}_{\Mav}(Q^2;T)=M_{\fd} M_{\fu}+ \alpha (1-\alpha)Q^2, 
\eea
and,
\bea
 \mathcal{C}_1^\mathcal{T}(z;T)\equiv z\,\bar{\mathcal{C}}_1^\mathcal{T}(z;T)\,.
\eea

The canonical normalization conditions for the longitudinal and transverse components can be written as
\bea
\nn \frac{1}{\left(E_{\Mav}^{(L,T)}\right)^2} = \frac{9m_G^2}{4\pi\alpha_{\rm IR}}\frac{d}{dQ^2}{\mathcal K}_{\Mav}^{(L,T)}(Q^2)\bigg|_{Q^2=-\left(m_{\Mav}^{(L,T)}\right)^2}\,,\\
\label{canonAV}
\eea
when the temperature dependence is omitted. At $T=0$ MeV, spin–orbit repulsion is introduced into the AV meson channel through a phenomenological coupling, $g_{\rm SO} \leq 1$. This accounts for the fact that dynamical chiral symmetry breaking generates a large dressed-quark anomalous chromomagnetic moment, which in turn significantly enhances the spin–orbit splitting between ground-state mesons and their parity partners~\cite{Bermudez:2017bpx,Bashir:2011dp,Chang:2010hb,Chang:2011ei}. For screening masses, however, it is evident that this factor, previously treated as a constant parameter, must now acquire a temperature dependence. Therefore, we adopt the phenomenological spin-orbit coupling as,
\bea
\label{gsomsc}
{\mathfrak g}_{\rm SO}^{q\bar{q},1^+}(T) = 1-\frac{M_u(T)}{M_u(0)}(1-{0.25}^2)\,,
\eea
such that at $T=0$ MeV, we obtain
\bea
{\mathfrak g}_{\rm SO}^{q\bar{q},1^+}(0) &= {0.25}^2\,,
\eea
which corresponds to the same value reported in Refs.~\cite{Lu:2017cln,Yin:2021uom} to determine the empirical inertial mass splittings of the $a_1$–$\rho$ and $\sigma$–$\rho$ systems. Explicitly, we introduce this spin-orbit effects in the determination of the screening masses by shifting the AV kernels in \eqn{eigenAV} as, 
\bea
\mathcal{K}_{\Mav}^{(L,T)}(Q^2;T) \to
{\mathfrak g}_{\rm SO}^{q\bar{q},1^+}(T) \,\, \mathcal{K}_{\Mav}^{(L,T)}(Q^2;T) \,.
\eea
With all these ingredients, the solution of \eqn{eigenAV} yields the screening masses of the AV mesons for the longitudinal and transverse modes, evaluated at $Q^2=-m_{\Mav}^2(T)$.
In Table \ref{table-mesones-av}, we present our results at $T=0$ MeV for ten mesons. Whenever experimental data are available, we provide a comparison to assess the accuracy of our predictions. At zero temperature, our results are consistent with the experimental findings \cite{ParticleDataGroup:2024cfk}, demonstrating that the model captures the essential features of the meson spectrum in this regime. \\
\begin{table}[b]
\caption{\justifying \label{table-mesones-av}
Computed masses for AV mesons (GeV) and BSAs with the parameters of Table~\ref{parameters} and Table~\ref{table-M}. }
\begin{center}
\begin{tabular}{@{\extracolsep{0.3 cm}}ccccc}
\hline
\hline
Meson   &Exp.& CI & $E_{\Mav}$ & $2M_R$ \,\\ \hline 
\rule{0ex}{2.5ex}
$ a_1(\tu\bar{\td})$ & 1.26 & 1.37 & 0.32 & 0.367   \\
\rule{0ex}{2.5ex}
$K_1^*(\tu\bar{\ts})$  & 1.34 & 1.48 & 0.32 & 0.433 \\
\rule{0ex}{2.5ex}
$f_1(\ts\bar{\ts})$ & 1.43 & 1.58 & 0.31 & 0.53  \\
\rule{0ex}{2.5ex}
$D_1(\tc\bar{\tu})$ & 2.42 &  2.39 &  0.20 &  0.591   \\
\rule{0ex}{2.5ex}
$D_{\ts 1}(\tc\bar{\ts})$ & 2.46 &  2.49 &  0.19 &  0.785 \\
\rule{0ex}{2.5ex}
$B_1(\tu\bar{\tb})$ & 5.72 & 5.54 & 0.10 & 0.681   \\
\rule{0ex}{2.5ex}
$B_{\ts 1}(\ts\bar{\tb})$ & 5.83 & 5.64 & 0.10 & 0.953   \\
\rule{0ex}{2.5ex}
$B_{\tc\tb}(\tc\bar{\tb})$ & $\cdots$  & 6.46 & 0.04 & 2.30 \\
\rule{0ex}{2.5ex}
$\chi_{\tc 1}(\tc\bar{\tc})$ & 3.51 & 3.37 & 0.08 & 1.52   \\
\rule{0ex}{2.5ex}
$\chi_{\tb 1}(\tb\bar{\tb})$ & 9.89 & 9.51 & 0.02 & 4.75   \\
\hline
\hline
\end{tabular}
\end{center}
\end{table}
Figure \ref{fig:av} illustrates the temperature dependence of the screening masses over the range $T/T_c\in[0,250]$ MeV.
Transverse results are shown with solid lines, longitudinal results with dashed lines. Analogously to the vector mesons, we show the free-theory limit with a black dashed line in the case of light mesons. Our results are compared with lattice data \cite{Bazavov:2020teh,Cheng:2010fe,Mukherjee:2008tr}, an effective chiral Lagrangian \cite{Song:1993af}, and QCD sum-rule predictions~\cite{Yazici:2016foi}. It should be stressed that the results derived from the last two approaches are reliable only for temperatures lower than the critical temperature, for the same reasons explained for V mesons.\\
The transverse screening masses of the lightest mesons increase monotonically up to $T_H$, then decrease to a minimum before rising again, while the longitudinal modes show a continuous decrease. In contrast, heavy-quark mesons display a monotonically increasing trend throughout the temperature range. The corresponding values of $T_H$ and the minima of each transverse mode of AV mesons are summarized in Table \ref{temps-avec}. It is evident that the $B_{\ts 1}$, $B_{\tc\tb}$, $\chi_{\tc 1}$, and $\chi_{\tb 1}$ mesons do not exhibit a minimum and remain monotonically increasing; in this case, the values of $T_H$ are determined using Eq.~(\ref{th}).
\\
\begin{table}[b]
 \caption{ \justifying \label{temps-avec} 
$T_H$ for the AV mesons in the transverse mode, indicating the point where the screening mass curve begins to decrease. Values in parentheses denote the minima of each curve, where present. The values marked with $^{\star}$ are obtained using Eq.~(\ref{th}).} 
\begin{center}
\label{temp2}
\begin{tabular}{@{\extracolsep{0.0 cm}}  c | c | c | c }
\hline \hline
 \, Meson \, &\,  $T_{H}$ \, & \, Meson \, &\,  $T_{H}$ \, 
 \\
 \hline
 \rule{0ex}{2.5ex}
$\, a_1 (\tu\overline{\td})$ & 0.69 (1.49)& $\, B_1(\tu\overline{\tb}) $ &\, 0.82 (1.46) \, \\ 
\rule{0ex}{2.5ex}
$\,  K_1^*(\tu\overline{\ts}) $ & \, 0.75 (1.52)\, & $\, B_{\ts 1}(\ts\overline{\tb}) $ & 0.37$^{\star}$ \\ 
\rule{0ex}{2.5ex}
$\,  f_1(\ts\overline{\ts}) $ & \, 0.89 (1.56)\, & $\, B_{\tc\tb}(\tc\overline{\tb}) $ & 0.39$^{\star}$\\ 
\rule{0ex}{2.5ex}
$\,   D_1(\tc\overline{\tu})$ & \, 0.83 (1.47)\, & $\, \chi_{\tc 1}(\tc\overline{\tc}) $ &  0.42$^{\star}$\\ 
\rule{0ex}{2.5ex}
$\,  D_{\ts 1}(\tc\overline{\ts})$   & 1.30 (1.45)  & $\, \chi_{\tb 1}(\tb\overline{\tb}) $ &  0.41$^{\star}$\\
\hline \hline
\end{tabular}
\end{center}
\end{table}
In addition, our results reveal the following hierarchy among the 
$T_H$  temperatures,
\bea
\nn &&T_H^{a_1} <T_H^{K_1^*}<T_H^{f_1}\,,\\
\nn &&T_H^{D_1} <T_H^{D_{\ts1}}\,,\\
&& T_H^{B_{\ts1}}<T_H^{B_{\tc\tb}}<T_H^{\chi_{\tb 1}} \,.
\eea
In most cases, our results align closely with those from other models. However, the limited availability of data for mesons with heavy quarks often leads various models using different approaches to reach differing conclusions.
Nevertheless, at very high temperatures ($T\to \infty$), these masses are expected to approach the value $2\sqrt{(\pi T)^2 +(2 M_R)^2 }$, independently of spin and flavor~\cite{Detar:1987kae}. 
In Table \ref{table-mesones-desviaciones}, we show the deviation from this limit at $T=500$ MeV using our results.
\begin{table}[h!]
\caption{\justifying \label{table-mesones-desviaciones}
Screening masses in GeV at $T=500$ MeV, showing their deviation from the $2\sqrt{(\pi T)^2 +(2 M_R)^2 }$ limit at this temperature.}
\begin{center}
\begin{tabular}{lcc|lcc}
\toprule
\multicolumn{3}{c|}{Vector} & \multicolumn{3}{c}{Axial-Vector} \\
\hline
  & Mass  &\, Dev (\%) \, &  &  Mass  & \, Dev (\%) \, \\
  \hline
\rule{0ex}{2.5ex}
$\rho(\tu\bar{\td})$ & 2.450 & 27.00  & $\, a_1(\tu\overline{\td})$ & 2.457 & 27.34 \\
\rule{0ex}{2.5ex}
$K_1(\tu\bar{\ts})$ & 2.463 & 27.50& $\,  K_1^*(\tu\overline{\ts}) $ & 2.470 &27.81\\
\rule{0ex}{2.5ex}
$\phi(\ts\bar{\ts})$ & 2.472 & 28.60 & $\, f_1(\ts\overline{\ts}) $ & 2.487 & 29.15   \\
\rule{0ex}{2.5ex}
$D^{*0}(\tc\bar{\tu})$ & 3.162 & 5.65 &$\,   D_1(\tc\overline{\tu})$ & 3.172 & 5.97 \\
\rule{0ex}{2.5ex}
$D_{\ts}^*(\tc\bar{\ts})$ & 3.143 & 8.83 &$\,  D_{\ts 1}(\tc\overline{\ts})$ & 3.215 & 11.09 \\
\rule{0ex}{2.5ex}
$B^{+*}(\tu\bar{\tb})$ & 6.333 &59.75 &$\, B_1(\tu\overline{\tb})$ & 6.342 & 59.72 \\
\rule{0ex}{2.5ex}
$B_{\ts}^{0*}(\ts\bar{\tb})$ & 6.311 & 53.82 &$\, B_{\ts 1}(\ts\overline{\tb})$ & 6.381 & 52.80 \\
\rule{0ex}{2.5ex}
$B_{\tc}^*(\tc\bar{\tb})$ & 7.028 &24.65&$\, B_{\tc\tb}(\tc\overline{\tb})$ & 7.137 & 23.13 \\
\rule{0ex}{2.5ex}
$\Jpsi(\tc\bar{\tc})$ & 3.912 &3.62 &$\, \chi_{\tc 1}(\tc\overline{\tc}) $ & 4.216 & 11.09\\
\rule{0ex}{2.5ex}
$\Upsilon(\tb\bar{\tb})$ & 9.853 &0.77 &$\, \chi_{\tb 1}(\tb\overline{\tb}) $ & 9.929 & 1.54  \\
\hline
\hline
\end{tabular}
\end{center}
\end{table}

\begin{figure*}[t]
\centering
\begin{subfigure}[b]{0.47\linewidth}
 \centering
\includegraphics[width=\linewidth]{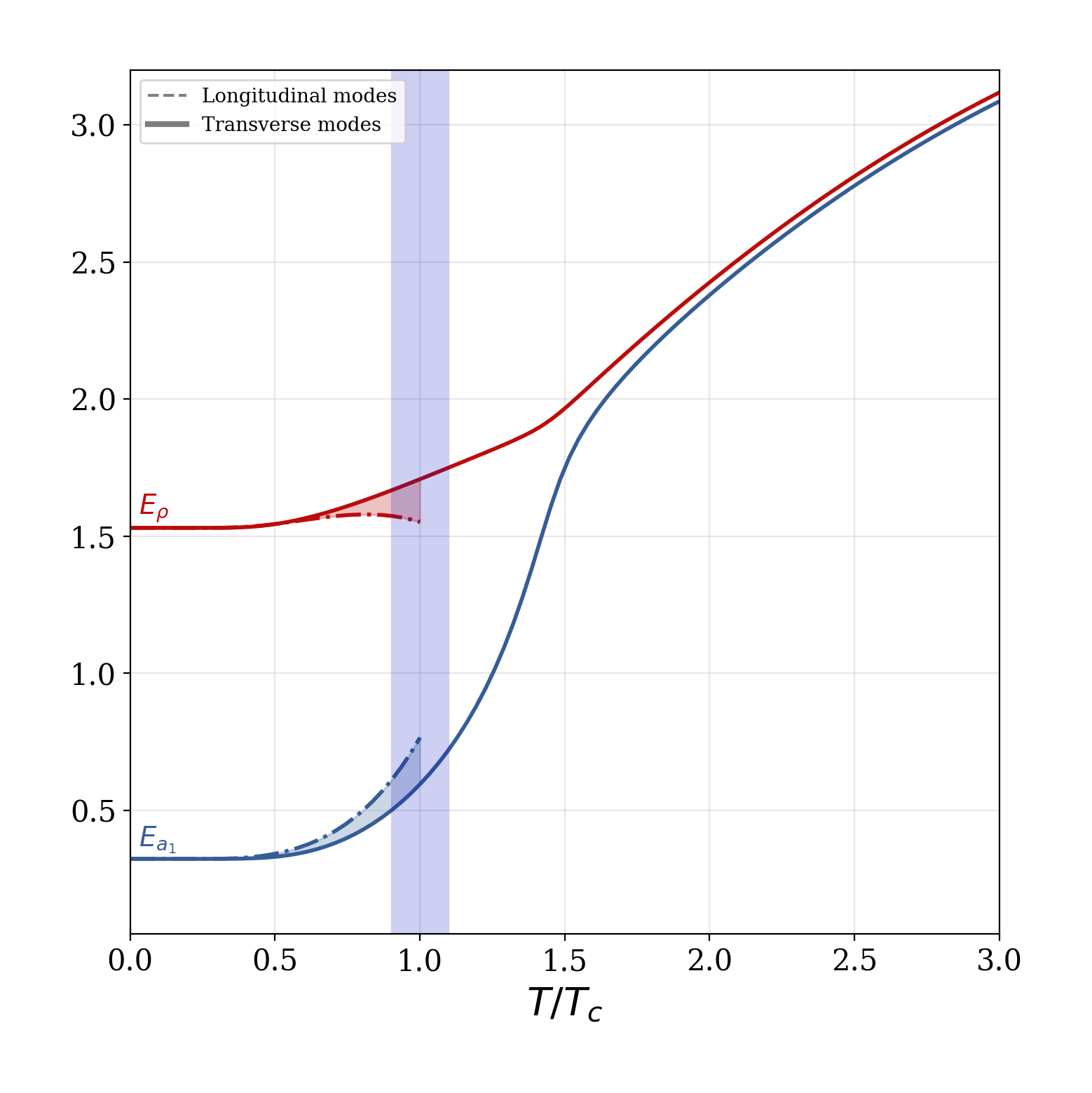}
\caption{BSAs for the lightest mesons}
\label{fig:bsamesligeros}
\end{subfigure}
\begin{subfigure}[b]{0.47\linewidth}
 \centering
\includegraphics[width=\linewidth]{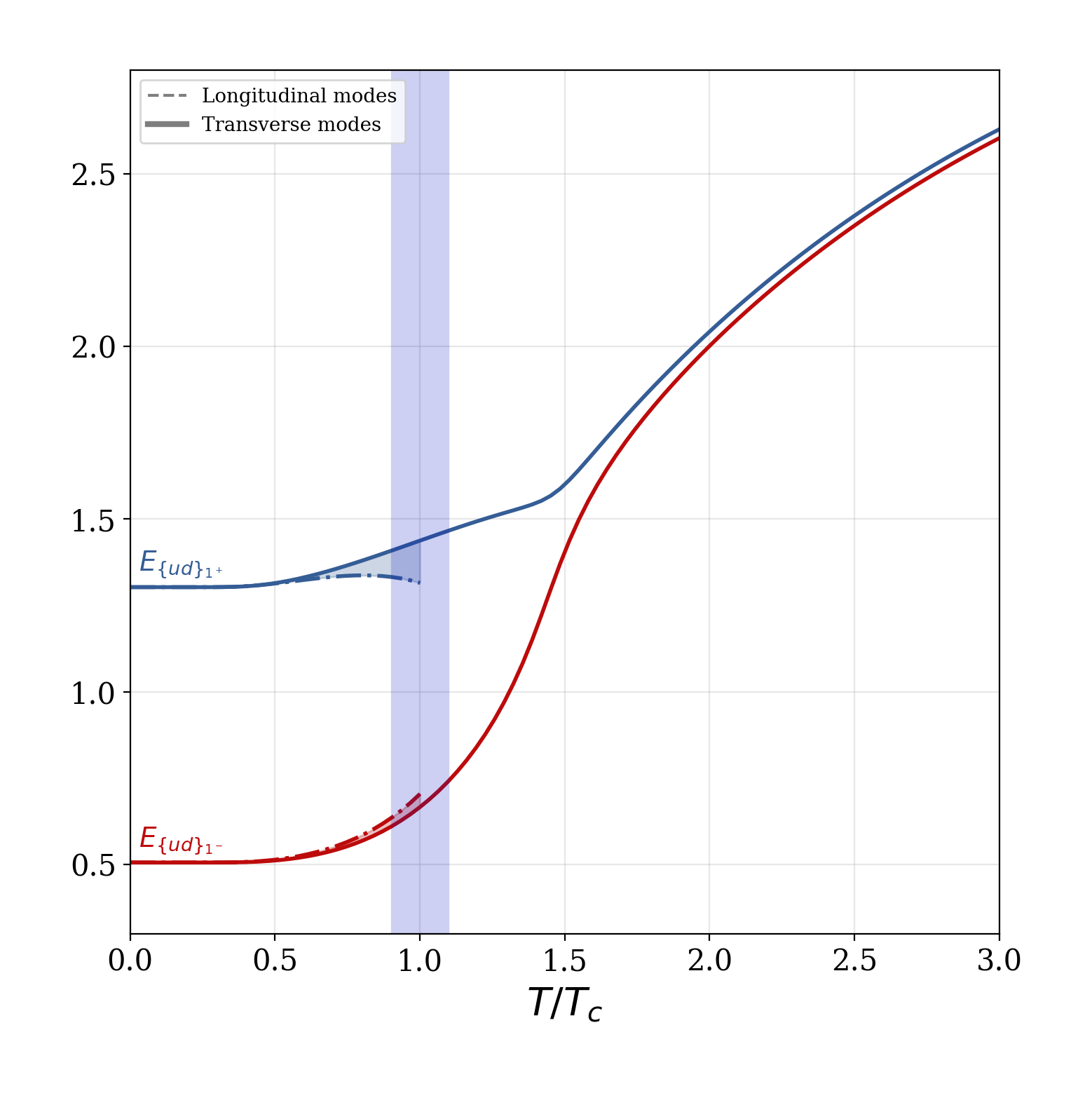}
\caption{BSAs for the lightest diquarks}
\label{fig:bsadiqligeros}
\end{subfigure}
\caption{\justifying Transverse and longitudinal components of the BSAs for the V and AV mesons, as well as for the corresponding diquark channels, evaluated at $Q^{\mu} = (\vec{Q}, \omega_0)$ and shown as functions of $T$. The plot highlights the convergence of $E_{\rho}$ and $E_{a_1}$ toward a common value at high temperatures. The shaded purple band denotes the critical region $T_c \pm 0.1\, T_c$. A similar trend is observed for the parity-partner diquarks, whose amplitudes also become nearly degenerate at large temperatures.
 }
\label{fig:ampvav}
\end{figure*}

The largest deviation, over 59\%, is observed for mesons composed of the lightest and the heaviest quarks, $\tu$ and $\tb$, while the smallest difference, below 2\%, occurs for the heaviest mesons composed of two $\tb$ quarks.
The results obtained at $T=500$ MeV allow us to observe the following mass splittings between mesons
\bea
\nn m_{\rho}(\tu\bar{\td})-  m_{a_1}(\tu\overline{\td})&=&  -0.007  \, {\rm GeV},\\
\nn m_{K_1}(\tu\bar{\ts})-m_{K_1^*}(\tu\overline{\ts}) &=&-0.007  \,{\rm MeV},\\
\nn m_{\phi}(\ts\bar{\ts})-m_{f_1}(\ts\overline{\ts})&=&  -0.015  \,{\rm GeV},\\
\nn m_{D^{*0}}(\tc\bar{\tu})- m_{D_1}(\tc\overline{\tu}) &=&  -0.01 \,{\rm GeV},\\
\nn m_{D_{\ts}^*}(\tc\bar{\ts}) -m_{D_{\ts 1}}(\tc\overline{\ts})&=&  -0.072, \,{\rm GeV},\\
\nn m_{B^{+*}}(\tu\bar{\tb})- m_{B_1}(\tu\overline{\tb}) &=&  -0.009 \,{\rm GeV},\\
\nn m_{B_{\ts}^{0*}}(\ts\bar{\tb})-m_{B_{\ts 1}}(\ts\overline{\tb})&=& -0.07  \,{\rm GeV},\\
\nn m_{B_{\tc}^*}(\tc\bar{\tb})-\, m_{B_{\tc\tb}}(\tc\overline{\tb}) &=&-0.109  \,{\rm GeV},\\
\nn m_{\Jpsi}(\tc\bar{\tc}) - m_{\chi_{\tc 1}}(\tc\overline{\tc}) &=&-0.304 \,{\rm GeV},\\
m_{\Upsilon}(\tb\bar{\tb}) -m_{ \chi_{\tb 1}}(\tb\overline{\tb}) &=&-0.076 \,{\rm GeV}.
\eea
From these equations, it is straightforward to see that the masses of the parity partners are nearly equal at this temperature. For example, in the light sector, the mass differences were approximately 750 MeV at zero temperature; however, at $T=500$ MeV the differences are reduced to less than 2\%.\\
On the other hand, the bound-state amplitudes are obtained from the homogeneous BSE. Their temperature dependence, for both longitudinal and transverse modes, is shown in Fig.~\ref{fig:ampvav} for the lightest quark systems in the V and AV channels.
 From Fig.~\ref{fig:ampvav}, several key features can be identified. At $T=0$ MeV, the amplitudes correspond to meson states previously reported in the literature and are consistent with experimental values. At this temperature, the longitudinal and transverse BSAs coincide, as expected for the ground-state configurations. As with the screening masses, the longitudinal BSAs can only be reliably extracted at temperatures below $T_c$.
 However, for the transverse component, which is considerably more stable, we observe that as temperature increases, the behavior of the amplitudes closely mirrors that of the screening masses in both the V and AV channels. Above $2.5\,T_c$, the difference between the amplitudes $E_{a_1}$ and $E_{\rho}$ 
  drops below 1.23\%, signaling that the lightest opposite-parity meson channels become nearly degenerate, which is a clear indication of the onset of symmetry restoration. Figure \ref{fig:ampvav}  also shows the BSAs for diquarks, which are analyzed in the following subsection.
\subsection{Diquarks}
\label{Diquarks-s}
\begin{figure*}[ht]
\begin{tabular}{@{\extracolsep{-2.3 cm}}c}
 \renewcommand{\arraystretch}{-1.6} %
 \hspace{-1cm}
 \includegraphics[scale=0.5]{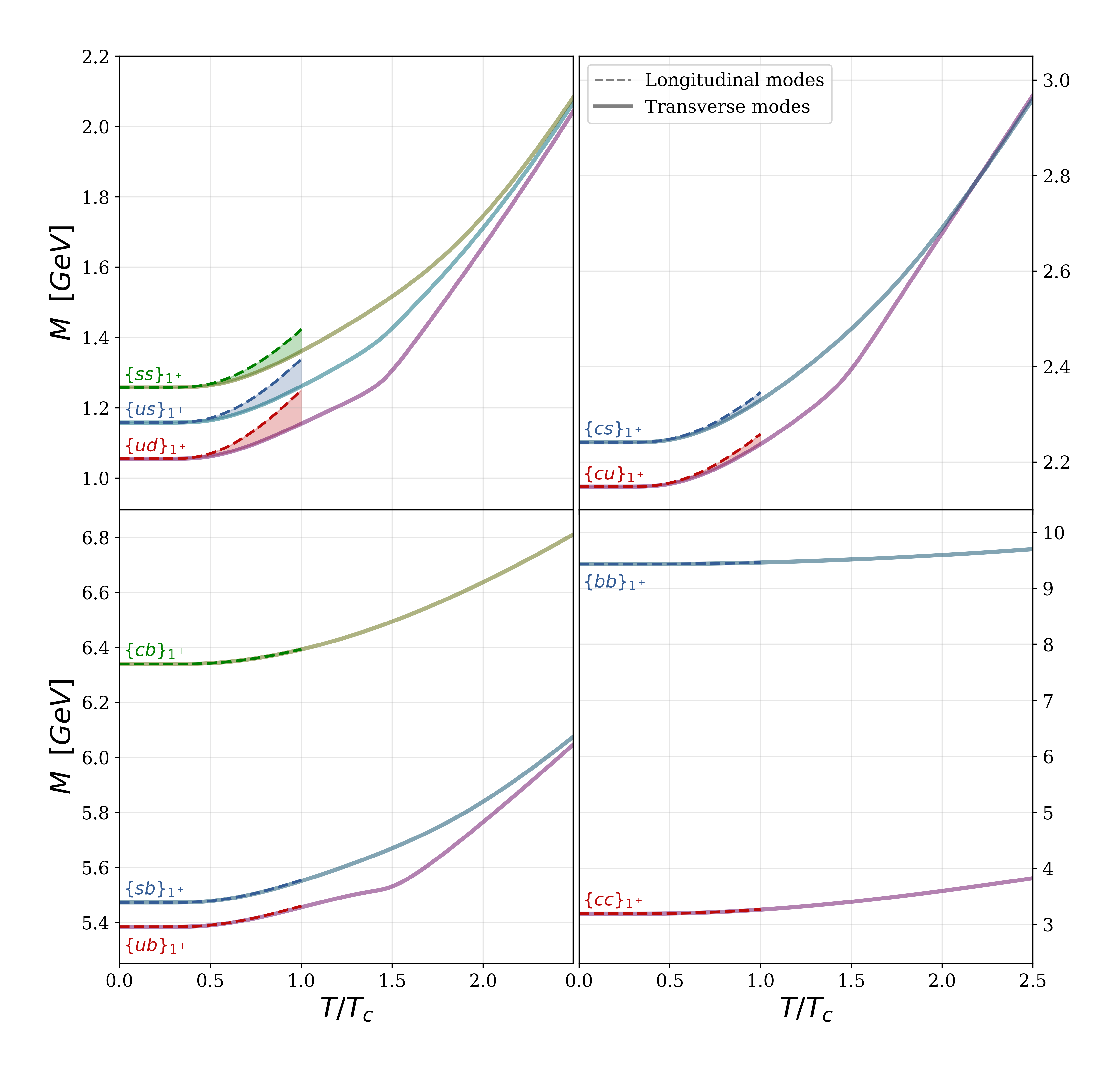}
\end{tabular}
\vspace{-1cm}
\caption{\justifying \label{plotdiav} Longitudinal and transverse screening masses for AV diquarks composed of the five quark flavors ($\tu$, $\td$, $\ts$, $\tc$ and $\tb$).The longitudinal components are shown with dashed lines, while the transverse components are shown with solid lines. Shaded areas elucidate differences between longitudinal and transverse modes.}
\end{figure*}
\begin{figure*}[ht]
\begin{tabular}{@{\extracolsep{-2.3 cm}}c}
 \renewcommand{\arraystretch}{-1.6} %
 \hspace{-1cm}
 \includegraphics[scale=0.5]{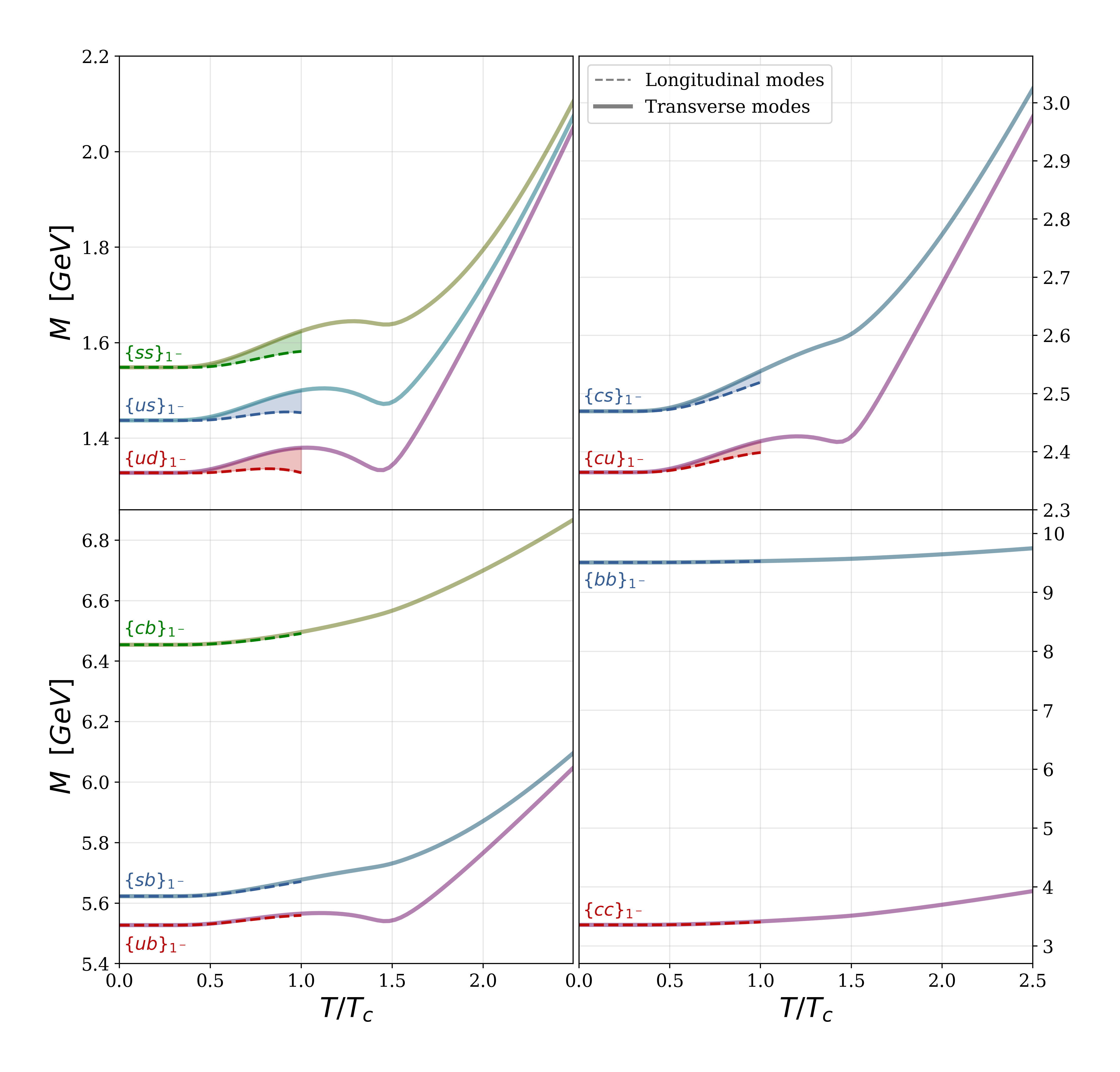}
\end{tabular}
\vspace{-1cm}
\caption{\justifying \label{plotdi} Longitudinal and transverse screening masses for V diquarks composed of the five quark flavors($\tu$, $\td$, $\ts$, $\tc$ and $\tb$).The longitudinal components are shown with dashed lines, while the transverse components are shown with solid lines. Shaded areas elucidate differences between longitudinal and transverse modes.}
\end{figure*}
\begin{figure*}[ht]
\centering
\begin{subfigure}[b]{0.47\linewidth}
 \centering
\includegraphics[width=\linewidth]{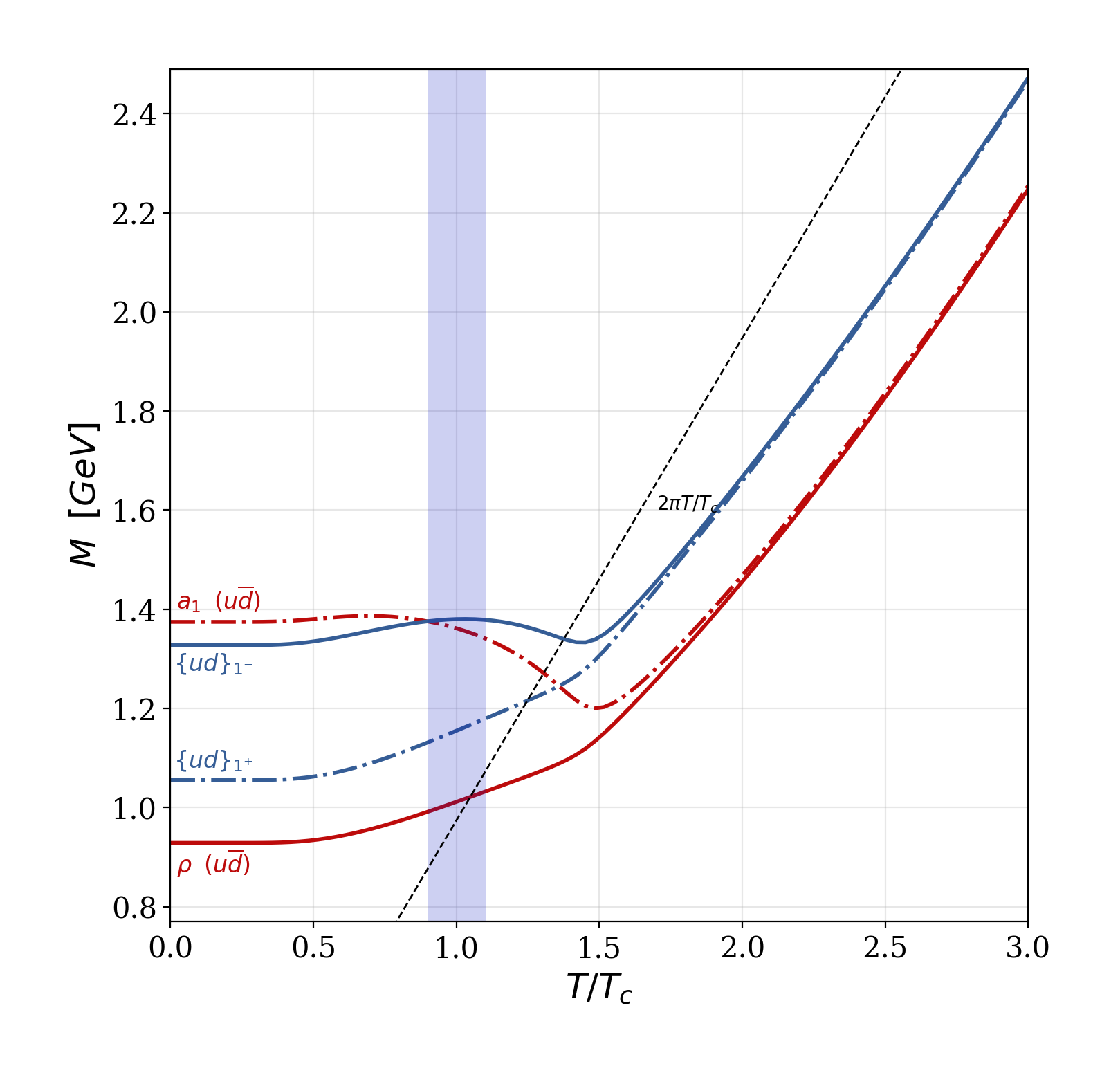}
\caption{Lightest mesons and diquarks}
\label{fig:ligeros}
\end{subfigure}
\begin{subfigure}[b]{0.474\linewidth}
 \centering
\includegraphics[width=\linewidth]{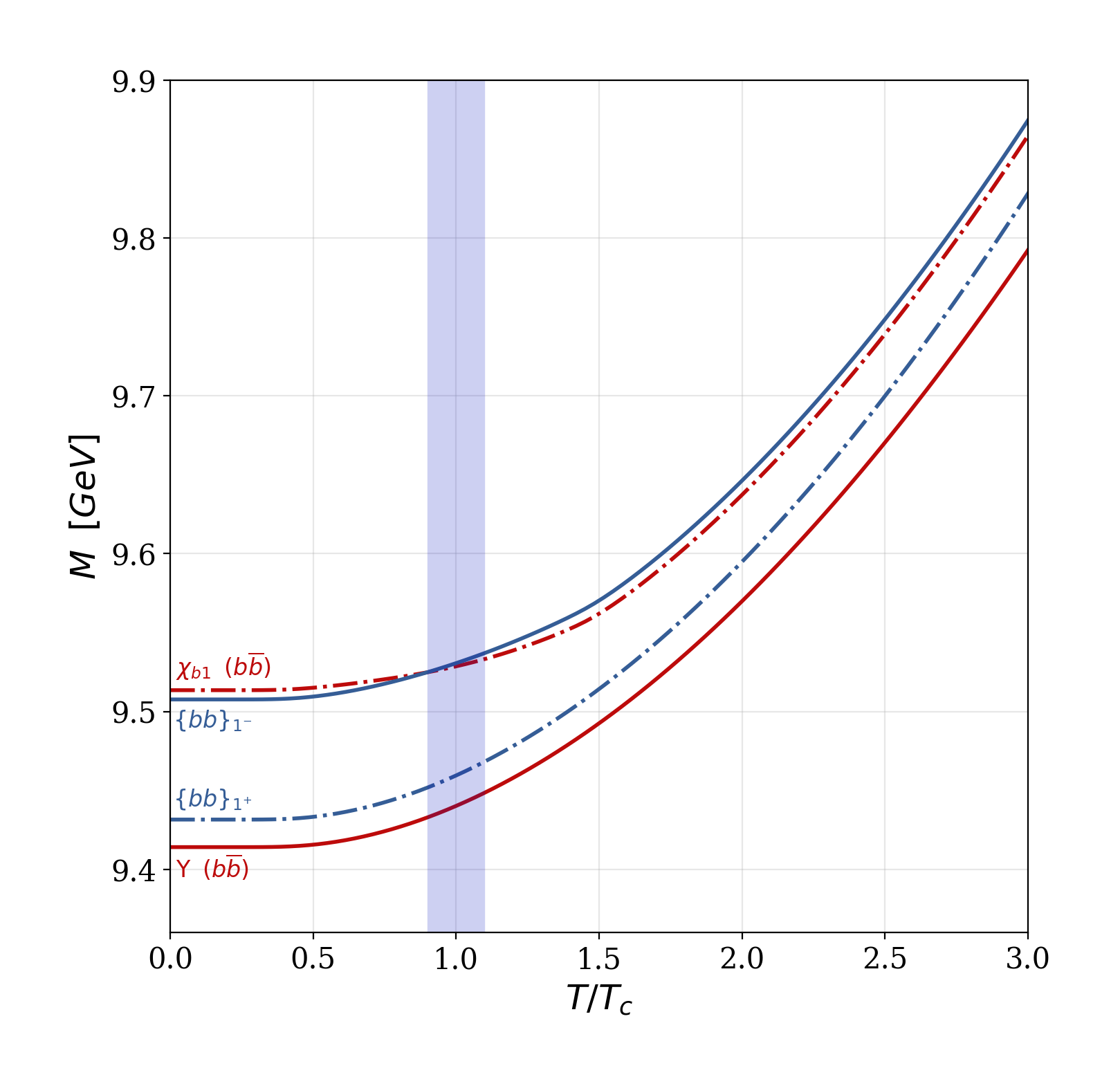}
\caption{Heaviest mesons and diquarks}
\label{fig:pesados}
\end{subfigure}
\caption{\justifying Comparison of the screening masses for the lightest and heaviest mesons and diquarks in the V and AV channels. Red lines denote mesonic states, while blue lines correspond to diquark states; dash-dotted curves represent positive-parity channels, and solid curves represent negative-parity channels. The black dashed line corresponds to the free-theory limit, $m = 2\pi T/T_c$.}
\label{fig:diyme}
\end{figure*}

Diquarks play a central role in the description of baryons within the quark–diquark framework, as a diquark in a color–antitriplet configuration can combine with a quark to form a color–singlet baryon. Such configurations are commonly referred to as good diquarks~\cite{Wilczek:2004im}. In this study, we restrict our attention to $qq$ states in a color–antitriplet representation and establish a reference for their screening masses, which are related to those of mesons partners, Here, we focus exclusively on the spin–1 diquark partners, whose meson partners are 
\begin{table}[H]
    \centering
    \begin{tabular}{ccc}
     Meson & & Diquark Partner\\
     \hline \hline
      \rule{0ex}{2.5ex}
   Meson vectorial  & $\to$ &  Diquark Axial Vector\\
   \rule{0ex}{2.5ex}
    Meson Axial Vector & $\to$ & Diquark Vectorial
    \end{tabular}
    \label{tab:placeholder}
\end{table}
Once the meson equations have been solved, the corresponding diquark partners follow immediately, since we only need to multiply the kernel by a factor of one half, namely,
\bea
\nn  {\cal K}_{\Me} &\to&   \frac{1}{2} {\cal K}_{\Me} \text{ in \eqn{bsevc}} \, ,\\
    \rule{0ex}{2.5ex}
  {\cal K}_{\Mav}  &\to&  \frac{1}{2} {\cal K}_{\Mav} \text{ in \eqn{eigenAV}}\, .
    \label{tab:placeholder}
\eea
The explicit eigenvalue equation for the AV diquark masses, $m^{(L,T)}_{\Dav}$, is
\bea
\label{bsevcd}
 1-\frac{1}{2}{\mathcal K}_{\Me}^{(L,T)}\left(Q^2=-\left(m^{(L,T)}_{\Dav}\right)^2;T\right) &=&0\,,
\eea
with, 
\bea
\nn \frac{1}{\left(E_{\Dav}^{(L,T)}\right)^2} = \frac{3m_G^2}{2\pi\alpha_{\rm IR}}\frac{d}{dQ^2}{\mathcal K}_{\Me}^{(L,T)}(Q^2)\bigg|_{Q^2=-\left(m_{\Dav}^{(L,T)}\right)^2}\,, \\
\label{canovcDav}
\eea
the canonical normalisation condition when omitting the temperature dependence.\\

Similarly, for V diquark masses, $m_{\De}^{(L,T)}$ , one obtains the following eigenvalue equation,
\bea
\label{bseavc}
 1+\frac{1}{2}{\mathcal K}_{\Mav}^{(L,T)}\left(Q^2=-\left(m_{\De}^{(L,T)}\right)^2;T\right) &=&0\,,
\eea
and 
\bea
\label{eqncdv}
\nn \frac{1}{\left(E_{\De}^{(L,T)}\right)^2} = \frac{3m_G^2}{2\pi\alpha_{\rm IR}}\frac{d}{dQ^2}{\mathcal K}_{\Mav}^{(L,T)}(Q^2)\bigg|_{Q^2=-\left(m_{\De}^{(L,T)}\right)^2}\,, \\
\label{canovcDv}
\eea
is the canonical normalisation condition. In  \eqn{eqncdv} the dependence with the temperature is understood.\\

In Table~\ref{fbod}, we present the results for the diquarks obtained at $T=0$ MeV, which are in excellent agreement with those reported in Ref.~\cite{Gutierrez-Guerrero:2021rsx} and were used to calculate baryon masses. This fact implies that the computation of baryons at finite temperature within the quark-diquark framework will require the inputs calculated in this work.
\begin{table}[t] %
    \centering
    \caption{\justifying \label{fbod} Masses at $T=0$ MeV for V diquarks ($DV$) and AV diquarks ($DAV$) obtained using the parameters described in Tab. \ref{parameters} and Tab. \ref{table-M}.}
    \begin{tabular}{@{\extracolsep{0.1 cm}} ccc|ccccc|ccc|ccccc}
         \hline
        \hline
        & & & & $DV$ & $DAV$  & &  & & &  & &  $DV$ && $DAV$ & \\  
         \hline
       \multirow{1}{*} & {$\tu\td$} &  & &1.32 & 1.05 & & & & {$\tu\tb$} && & 5.52 && 5.38 & \\    
        \rule{0ex}{2.5ex}
         \multirow{1}{*} & {$\tu\ts$} &  && 1.43 & 1.15 & &&& {$\ts\tb$} &&& 5.62 && 5.47 & \\
          \rule{0ex}{2.5ex}
         \multirow{1}{*}& {$\ts\ts$}  &  && 1.54 & 1.25 & &&& {$\tc\tb$} &&& 6.45 && 6.33 & \\
          \rule{0ex}{2.5ex}
        \multirow{1}{*}& {$\tc\tu$} & && 2.36 & 2.14 & &&&{$\tc\tc$} &&& 3.35 && 3.19 &\\
         \rule{0ex}{2.5ex}
         \multirow{1}{*}& {$\tc\ts$} &  && 2.46 & 2.24 & &&& {$\tb\tb$} &&& 9.50 && 9.43  &\\
        \hline
        \hline
    \end{tabular}
\end{table}
Using the equations described above Eqs. (\ref{bsevcd}), we present our results for AV diquarks in Fig.~\ref{plotdiav}, displayed in both longitudinal and transverse modes. For clarity, a shaded area between the curves is included to emphasize their differences. It is readily seen that their behavior resembles that of the V mesons discussed in Sec.~\ref{Vector-M}, namely, they exhibit a monotonic increase.
This behavior has important implications for states that can be viewed as being composed of diquarks, since the contributions of these correlations cannot be neglected at high temperatures. The same applies to scalar diquarks  Ref.\cite{Ramirez-Garrido:2025rsu}, whose role must also be taken into account—for instance, in the calculation of baryon properties.
Our results for V diquarks are presented in Fig.~\ref{plotdi}. As expected, their behavior closely resembles that of AV mesons: they increase monotonically at low temperatures, then decrease until reaching a minimum, after which they rise again.\\
The BSAs for the lightest diquarks as functions of temperature are presented in Fig.~\ref{fig:ampvav}. A clear pattern emerges: the parity partners progressively approach one another as the temperature increases, mirroring the behavior observed in the corresponding mesonic channels.

A comparison between meson masses and their diquark counterparts in the light and heavy sector is shown in Fig.~\ref{fig:diyme}. This figure highlights not only the similarities in their behavior but also that diquarks with different parity become degenerate at high temperatures. Specifically, one finds
\begin{eqnarray}
\begin{array}{cc}
m_{a_1}\approx m_{\rho}, & {\mbox{ when }\,\,\,\,} T>1.67 \,\, T_c\, ,  \\
m_{\{ud\}_{1-}}\approx m_{\{ud\}_{1+}}, & {\mbox{ when }\,\,\,\,} T> 1.55 \,\, T_c \, .
\end{array}
\end{eqnarray}
In our study, for $T>1.67\,T_c$ the mass difference between the $a_1$ and $\rho$ mesons falls below 3\%, indicating that the parity partners become effectively degenerate. This suggests that the dynamical chiral symmetry breaking effect weakens progressively as the temperature increases, and the chiral symmetry can be restored once the temperature reaches the critical value \cite{Mo:2010zza}. In the heavy sector, meson and diquark screening masses converge toward their asymptotic values.
\section{ SUMMARY AND PERSPECTIVE}
\label{Summary}

From the results obtained in this work, we can highlight the main findings for spin-1 mesons in a symmetry-preserving treatment of a vector–vector CI at nonzero temperature.
\begin{itemize}
    \item The breaking of 
$O(4)$  symmetry leads to the emergence of two screening modes, longitudinal and transverse, for V and AV mesons. At 
$T=0$, however, these modes remain degenerate in both channels, consistently reflecting the unbroken 
$O(4)$  symmetry in the vacuum.

\item At $T=0$ MeV, our results for the masses of V and AV mesons are in agreement with experimental data, as shown in Tables \ref{table-mesones-vec} and \ref{table-mesones-av}. 

\item Figures \ref{fig:vectorSM} and \ref{fig:av} present the temperature dependence of the screening masses for V and AV mesons in their longitudinal and transverse modes. In general, we find that, at low temperatures the vector and axial-vector screening masses remain close to the pole mass of the ground state.
As $T$ increases, dynamical chiral symmetry breaking weakens, the quark mass decreases, and the screening
masses rise. Near $T_c$, different trends may appear depending on the theoretical treatment, while at high
temperature both channels approach the free limit $M \to 2\sqrt{(\pi T)^2 +M_{R}^2 }$, signaling chiral-symmetry restoration. Thus, we find the expected phenomenological behavior of the screening masses of V and AV mesons.

\item In the longitudinal channel, we find that the Bethe--Salpeter equation becomes numerically unstable already slightly below the critical temperature. This behavior is fully consistent with earlier Dyson--Schwinger analyses and with lattice QCD simulations, both of which report that longitudinal modes become unreliable or difficult to extract near and above $T_c$.
\begin{figure*}[t!]
\centering
\begin{subfigure}[b]{0.47\linewidth}
 \centering
\includegraphics[angle=-90, width=1\linewidth]{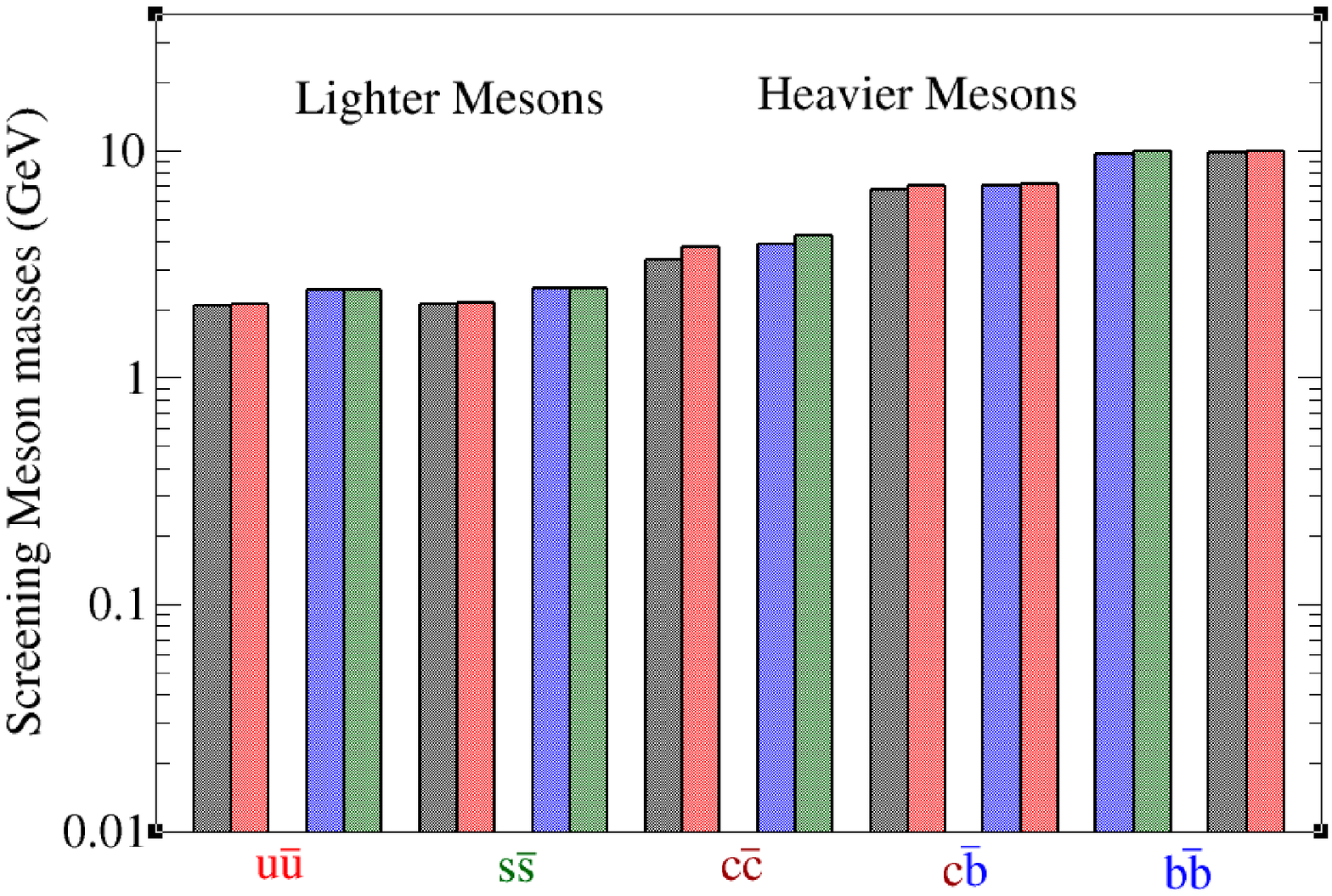}
\label{fig:ligeros}
\end{subfigure}
\begin{subfigure}[b]{0.474\linewidth}
 \centering
\includegraphics[angle=-90, width=1\linewidth]{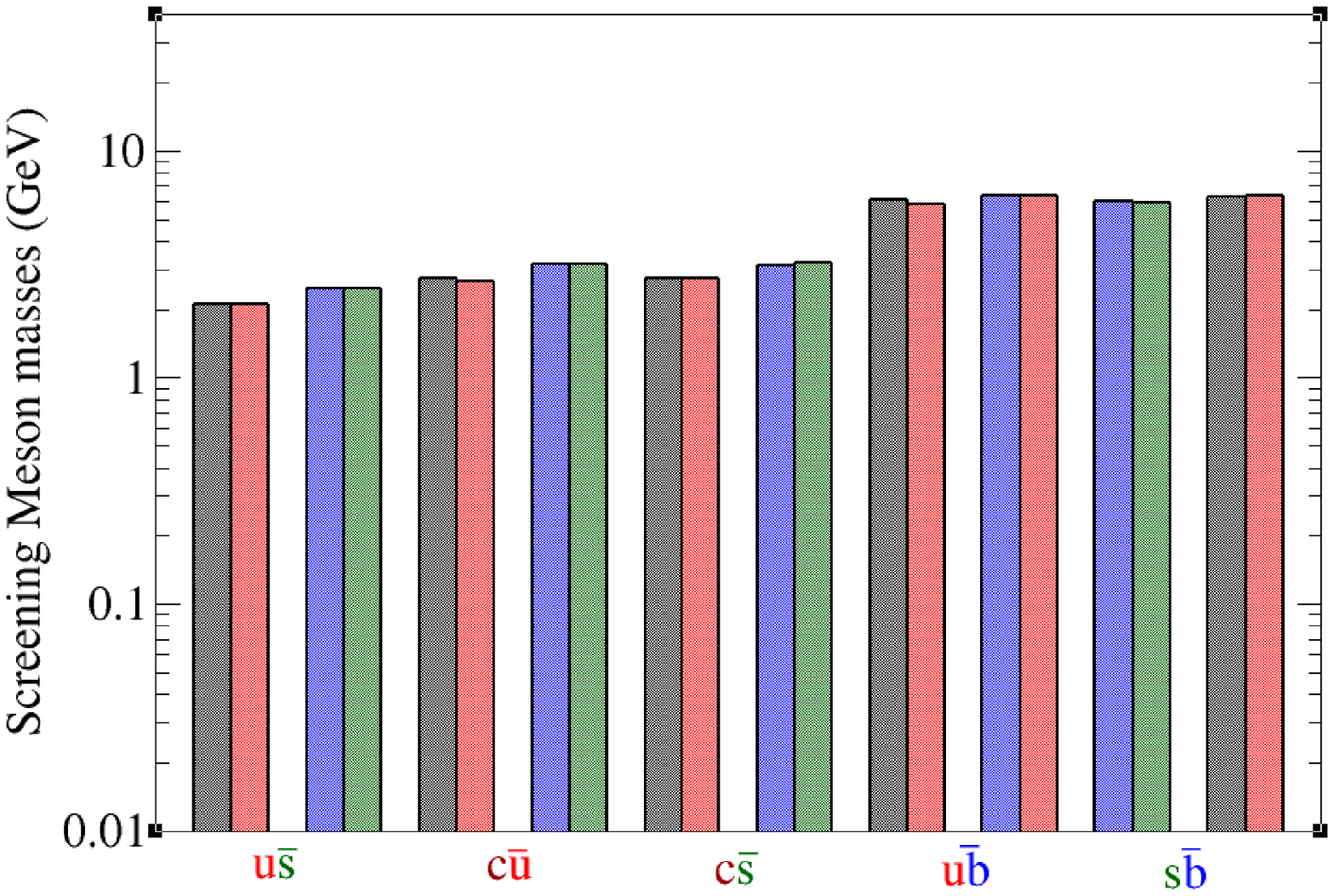}
\label{fig:pesados}
\end{subfigure}
\caption{\justifying Screening masses of scalar, pseudoscalar, vector and axial-vector mesons at $T=500$ MeV in red, black, blue and green columns respectively. On the l.h.s., the lighter and heavier mesons while in the r.h.s. the light and heavy-light mesons. Each meson is shown together with its corresponding chiral partner. 
}
\label{fig:parityS}
\end{figure*}

\item The analysis indicates that the transverse screening masses of V mesons exhibit a continuous increase, which accelerates above the temperature 
$T_H$. By contrast, AV mesons display a more intricate behavior, increasing up to 
$T_H$, then decreasing to a minimum, and subsequently rising again.
Our results show that the variation in behavior between V and AV is primarily due to the introduction of spin-orbit repulsion in the AV channel, implemented through the phenomenological coupling ${\mathfrak g}_{\rm SO}^{q\bar{q},1^+}(T)$.

\item Our results for the transverse screening masses of V and AV mesons were compared with those obtained using 
other approaches, showing good overall agreement. 
Nonetheless, although the CI framework predicts a smooth monotonic increase of the vector and axial-vector 
screening masses with temperature, several works in the literature report either decreasing or non-monotonic 
trends. Such discrepancies originate from the restricted validity of QCD sum rules, Refs.~\cite{Veliev:2011zz,Dominguez:2013fca}, and effective chiral 
Lagrangians, Ref.~\cite{Song:1993af}, which are only reliable for $T \leq T_c$ and rely on leading-order or hadronic-phase assumptions, thereby limiting their applicability near or beyond the transition point. Under contrast, 
the present CI treatment, applied uniformly to both light and heavy mesons, offers a consistent and 
complementary viewpoint that remains stable across the full temperature range considered.

\item The Bethe–Salpeter amplitudes of the lightest spin-1 mesons and diquarks at finite temperature are shown in Fig.~\ref{fig:ampvav}. One observes that the amplitudes of parity partners become nearly identical at high temperatures.

\item The transverse screening masses of the parity partners of the lightest mesons become nearly degenerate at high temperatures, as illustrated in Fig. \ref{fig:diyme} (a). For the heaviest mesons, see Fig. \ref{fig:diyme} (b), the differences are of the order 75 MeV and 45 MeV between mesonic $\tb\bar\tb$ and diquark $\tb\tb$ states respectively, both at $T=500$ MeV.

\item Our predictions for diquarks are shown in Fig. \ref{plotdi}, which displays their behavior in the longitudinal and transverse modes.

\item The behavior of AV diquarks is similar to that of V mesons, while V diquarks exhibit a behavior analogous to that of AV mesons, as expected.\\

\item We emphasize the deviations of the screening masses from the free limit, as summarized in Table \ref{table-mesones-desviaciones}.

\item Finally, combining the results obtained in this work with those of Ref. \cite{Ramirez-Garrido:2025rsu}, we can conclude that, within this model, the masses of the parity partners of mesons become degenerate at high temperatures for both spin-zero and spin-one mesons, indicating chiral symmetry restoration, as illustrated in Fig.~\ref{fig:parityS}.
\end{itemize}



This work provides CI model predictions for the V and AV meson screening masses over an extended temperature range, offering analytical expressions that serve as a reliable baseline for future studies and refinements. Overall, the results provide a coherent and comprehensive description of meson behavior at finite temperature, highlighting the emergence of longitudinal and transverse modes, the degeneracy of parity partners at high temperatures, and the restoration of chiral symmetry within this model. These findings can also serve as inputs for future calculations of baryon screening masses within the quark–diquark picture.
\vspace*{2mm}

\begin{acknowledgements}
\vspace*{-2mm}
The authors would like to thank the referee for their valuable feedback. L.~X.~Gutiérrez-Guerrero acknowledges the Secretaría de Ciencia, Humanidades, Tecnología e Innovación (SECIHTI) for the support provided to her through the Investigadores e Investigadoras por México SECIHTI program.
L.~X.~Gutiérrez-Guerrero, R.~J.~Hernández-Pinto, and M.~A.~Ramírez-Garrido thank SECIHTI for the support received through Project CBF2023-2024-268, Física Hadronica en JLab: Descifrando la Estructura Interna de Mesones y Bariones, from the 2023-2024 frontier science call. M. A. P\'erez de Le\'on is supported by SECIHTI (M\'exico) through the {\it{Estancias Posdoctorales por M\'exico}} program.
The work of L.~X.~Gutiérrez-Guerrero and R.~J.~Hernández-Pinto is partly supported by SECIHTI (Mexico) through the \it{Sistema Nacional de Investigadores}.
\end{acknowledgements}
\bibliography{ccc-a}

\begin{thebibliography}{80}
\expandafter\ifx\csname natexlab\endcsname\relax\def\natexlab#1{#1}\fi
\expandafter\ifx\csname bibnamefont\endcsname\relax
  \def\bibnamefont#1{#1}\fi
\expandafter\ifx\csname bibfnamefont\endcsname\relax
  \def\bibfnamefont#1{#1}\fi
\expandafter\ifx\csname citenamefont\endcsname\relax
  \def\citenamefont#1{#1}\fi
\expandafter\ifx\csname url\endcsname\relax
  \def\url#1{\texttt{#1}}\fi
\expandafter\ifx\csname urlprefix\endcsname\relax\def\urlprefix{URL }\fi
\providecommand{\bibinfo}[2]{#2}
\providecommand{\eprint}[2][]{\url{#2}}

\bibitem[{\citenamefont{Chen et~al.}(2024)\citenamefont{Chen, Gao, and
  Qin}}]{Chen:2024emt}
\bibinfo{author}{\bibfnamefont{C.}~\bibnamefont{Chen}},
  \bibinfo{author}{\bibfnamefont{F.}~\bibnamefont{Gao}}, \bibnamefont{and}
  \bibinfo{author}{\bibfnamefont{S.-x.} \bibnamefont{Qin}}
  (\bibinfo{year}{2024}), \eprint{2412.15045}.

\bibitem[{\citenamefont{Mukherjee}(2009)}]{Mukherjee:2008tr}
\bibinfo{author}{\bibfnamefont{S.}~\bibnamefont{Mukherjee}},
  \bibinfo{journal}{Nucl. Phys. A} \textbf{\bibinfo{volume}{820}},
  \bibinfo{pages}{283C} (\bibinfo{year}{2009}), \eprint{0810.2906}.

\bibitem[{\citenamefont{Dosch and Narison}(1988)}]{Dosch:1988vt}
\bibinfo{author}{\bibfnamefont{H.~G.} \bibnamefont{Dosch}} \bibnamefont{and}
  \bibinfo{author}{\bibfnamefont{S.}~\bibnamefont{Narison}},
  \bibinfo{journal}{Phys. Lett. B} \textbf{\bibinfo{volume}{203}},
  \bibinfo{pages}{155} (\bibinfo{year}{1988}).

\bibitem[{\citenamefont{Ayala et~al.}(2012)\citenamefont{Ayala, Dominguez,
  Loewe, and Zhang}}]{Ayala:2012ch}
\bibinfo{author}{\bibfnamefont{A.}~\bibnamefont{Ayala}},
  \bibinfo{author}{\bibfnamefont{C.~A.} \bibnamefont{Dominguez}},
  \bibinfo{author}{\bibfnamefont{M.}~\bibnamefont{Loewe}}, \bibnamefont{and}
  \bibinfo{author}{\bibfnamefont{Y.}~\bibnamefont{Zhang}},
  \bibinfo{journal}{Phys. Rev. D} \textbf{\bibinfo{volume}{86}},
  \bibinfo{pages}{114036} (\bibinfo{year}{2012}), \eprint{1210.2588}.

\bibitem[{\citenamefont{Hatsuda et~al.}(1993)\citenamefont{Hatsuda, Koike, and
  Lee}}]{Hatsuda:1992bv}
\bibinfo{author}{\bibfnamefont{T.}~\bibnamefont{Hatsuda}},
  \bibinfo{author}{\bibfnamefont{Y.}~\bibnamefont{Koike}}, \bibnamefont{and}
  \bibinfo{author}{\bibfnamefont{S.-H.} \bibnamefont{Lee}},
  \bibinfo{journal}{Nucl. Phys. B} \textbf{\bibinfo{volume}{394}},
  \bibinfo{pages}{221} (\bibinfo{year}{1993}).

\bibitem[{\citenamefont{Wang and Wang}(2015)}]{Wang:2015ynf}
\bibinfo{author}{\bibfnamefont{Z.-B.} \bibnamefont{Wang}} \bibnamefont{and}
  \bibinfo{author}{\bibfnamefont{Z.-G.} \bibnamefont{Wang}},
  \bibinfo{journal}{Acta Phys. Polon. B} \textbf{\bibinfo{volume}{46}},
  \bibinfo{pages}{2467} (\bibinfo{year}{2015}).

\bibitem[{\citenamefont{Brambilla et~al.}(2011)}]{Brambilla:2010cs}
\bibinfo{author}{\bibfnamefont{N.}~\bibnamefont{Brambilla}}
  \bibnamefont{et~al.}, \bibinfo{journal}{Eur. Phys. J. C}
  \textbf{\bibinfo{volume}{71}}, \bibinfo{pages}{1534} (\bibinfo{year}{2011}),
  \eprint{1010.5827}.

\bibitem[{\citenamefont{Matsui and Satz}(1986)}]{Matsui:1986dk}
\bibinfo{author}{\bibfnamefont{T.}~\bibnamefont{Matsui}} \bibnamefont{and}
  \bibinfo{author}{\bibfnamefont{H.}~\bibnamefont{Satz}},
  \bibinfo{journal}{Phys. Lett. B} \textbf{\bibinfo{volume}{178}},
  \bibinfo{pages}{416} (\bibinfo{year}{1986}).

\bibitem[{\citenamefont{Bochkarev and Shaposhnikov}(1986)}]{Bochkarev:1985ex}
\bibinfo{author}{\bibfnamefont{A.~I.} \bibnamefont{Bochkarev}}
  \bibnamefont{and} \bibinfo{author}{\bibfnamefont{M.~E.}
  \bibnamefont{Shaposhnikov}}, \bibinfo{journal}{Nucl. Phys. B}
  \textbf{\bibinfo{volume}{268}}, \bibinfo{pages}{220} (\bibinfo{year}{1986}).

\bibitem[{\citenamefont{Song}(1996)}]{Song:1995ga}
\bibinfo{author}{\bibfnamefont{C.}~\bibnamefont{Song}}, \bibinfo{journal}{Phys.
  Rev. D} \textbf{\bibinfo{volume}{53}}, \bibinfo{pages}{3962}
  (\bibinfo{year}{1996}), \eprint{hep-ph/9501364}.

\bibitem[{\citenamefont{Pisarski}(1995)}]{Pisarski:1995xu}
\bibinfo{author}{\bibfnamefont{R.~D.} \bibnamefont{Pisarski}},
  \bibinfo{journal}{Phys. Rev. D} \textbf{\bibinfo{volume}{52}},
  \bibinfo{pages}{R3773} (\bibinfo{year}{1995}), \eprint{hep-ph/9503328}.

\bibitem[{\citenamefont{Hern{\'a}ndez et~al.}(2025)\citenamefont{Hern{\'a}ndez,
  Mart{\'\i}nez-S{\'a}nchez, and Zamora}}]{Hernandez:2025inu}
\bibinfo{author}{\bibfnamefont{L.~A.} \bibnamefont{Hern{\'a}ndez}},
  \bibinfo{author}{\bibfnamefont{J.~D.}
  \bibnamefont{Mart{\'\i}nez-S{\'a}nchez}}, \bibnamefont{and}
  \bibinfo{author}{\bibfnamefont{R.}~\bibnamefont{Zamora}},
  \bibinfo{journal}{Phys. Rev. D} \textbf{\bibinfo{volume}{111}},
  \bibinfo{pages}{096019} (\bibinfo{year}{2025}), \eprint{2502.08051}.

\bibitem[{\citenamefont{Gao and Ding}(2020)}]{Gao:2020hwo}
\bibinfo{author}{\bibfnamefont{F.}~\bibnamefont{Gao}} \bibnamefont{and}
  \bibinfo{author}{\bibfnamefont{M.}~\bibnamefont{Ding}},
  \bibinfo{journal}{Eur. Phys. J. C} \textbf{\bibinfo{volume}{80}},
  \bibinfo{pages}{1171} (\bibinfo{year}{2020}), \eprint{2006.05909}.

\bibitem[{\citenamefont{Fischer}(2019)}]{Fischer:2018sdj}
\bibinfo{author}{\bibfnamefont{C.~S.} \bibnamefont{Fischer}},
  \bibinfo{journal}{Prog. Part. Nucl. Phys.} \textbf{\bibinfo{volume}{105}},
  \bibinfo{pages}{1} (\bibinfo{year}{2019}), \eprint{1810.12938}.

\bibitem[{\citenamefont{Carlomagno and
  Izzo~Villafa{\~n}e}(2019)}]{Carlomagno:2019yvi}
\bibinfo{author}{\bibfnamefont{J.~P.} \bibnamefont{Carlomagno}}
  \bibnamefont{and} \bibinfo{author}{\bibfnamefont{M.~F.}
  \bibnamefont{Izzo~Villafa{\~n}e}}, \bibinfo{journal}{Phys. Rev. D}
  \textbf{\bibinfo{volume}{100}}, \bibinfo{pages}{076011}
  (\bibinfo{year}{2019}), \eprint{1906.04257}.

\bibitem[{\citenamefont{Sumit et~al.}(2023)\citenamefont{Sumit, Haque, and
  Patra}}]{Sumit:2023hjj}
\bibinfo{author}{\bibnamefont{Sumit}},
  \bibinfo{author}{\bibfnamefont{N.}~\bibnamefont{Haque}}, \bibnamefont{and}
  \bibinfo{author}{\bibfnamefont{B.~K.} \bibnamefont{Patra}},
  \bibinfo{journal}{Phys. Lett. B} \textbf{\bibinfo{volume}{845}},
  \bibinfo{pages}{138143} (\bibinfo{year}{2023}), \eprint{2305.08525}.

\bibitem[{\citenamefont{Cheng et~al.}(2011)}]{Cheng:2010fe}
\bibinfo{author}{\bibfnamefont{M.}~\bibnamefont{Cheng}} \bibnamefont{et~al.},
  \bibinfo{journal}{Eur. Phys. J. C} \textbf{\bibinfo{volume}{71}},
  \bibinfo{pages}{1564} (\bibinfo{year}{2011}), \eprint{1010.1216}.

\bibitem[{\citenamefont{Bazavov and Weber}(2021)}]{Bazavov:2020teh}
\bibinfo{author}{\bibfnamefont{A.}~\bibnamefont{Bazavov}} \bibnamefont{and}
  \bibinfo{author}{\bibfnamefont{J.~H.} \bibnamefont{Weber}},
  \bibinfo{journal}{Prog. Part. Nucl. Phys.} \textbf{\bibinfo{volume}{116}},
  \bibinfo{pages}{103823} (\bibinfo{year}{2021}), \eprint{2010.01873}.

\bibitem[{\citenamefont{Petreczky}(2009)}]{Petreczky:2009at}
\bibinfo{author}{\bibfnamefont{P.}~\bibnamefont{Petreczky}},
  \bibinfo{journal}{Nucl. Phys. A} \textbf{\bibinfo{volume}{830}},
  \bibinfo{pages}{11C} (\bibinfo{year}{2009}), \eprint{0908.1917}.

\bibitem[{\citenamefont{Dominguez
  et~al.}(2013{\natexlab{a}})\citenamefont{Dominguez, Loewe, and
  Zhang}}]{Dominguez:2012um}
\bibinfo{author}{\bibfnamefont{C.~A.} \bibnamefont{Dominguez}},
  \bibinfo{author}{\bibfnamefont{M.}~\bibnamefont{Loewe}}, \bibnamefont{and}
  \bibinfo{author}{\bibfnamefont{Y.}~\bibnamefont{Zhang}},
  \bibinfo{journal}{Nucl. Phys. B Proc. Suppl.} \textbf{\bibinfo{volume}{234}},
  \bibinfo{pages}{305} (\bibinfo{year}{2013}{\natexlab{a}}),
  \eprint{1212.2241}.

\bibitem[{\citenamefont{Dey et~al.}(1990)\citenamefont{Dey, Eletsky, and
  Ioffe}}]{Dey:1990ba}
\bibinfo{author}{\bibfnamefont{M.}~\bibnamefont{Dey}},
  \bibinfo{author}{\bibfnamefont{V.~L.} \bibnamefont{Eletsky}},
  \bibnamefont{and} \bibinfo{author}{\bibfnamefont{B.~L.} \bibnamefont{Ioffe}},
  \bibinfo{journal}{Phys. Lett. B} \textbf{\bibinfo{volume}{252}},
  \bibinfo{pages}{620} (\bibinfo{year}{1990}).

\bibitem[{\citenamefont{Mamedov and Taghiyeva}(2021)}]{Mamedov:2021dpv}
\bibinfo{author}{\bibfnamefont{S.}~\bibnamefont{Mamedov}} \bibnamefont{and}
  \bibinfo{author}{\bibfnamefont{S.}~\bibnamefont{Taghiyeva}},
  \bibinfo{journal}{Eur. Phys. J. C} \textbf{\bibinfo{volume}{81}},
  \bibinfo{pages}{1080} (\bibinfo{year}{2021}), \eprint{2108.13513}.

\bibitem[{\citenamefont{Dominguez et~al.}(2010)\citenamefont{Dominguez, Loewe,
  Rojas, and Zhang}}]{Dominguez:2010ve}
\bibinfo{author}{\bibfnamefont{C.~A.} \bibnamefont{Dominguez}},
  \bibinfo{author}{\bibfnamefont{M.}~\bibnamefont{Loewe}},
  \bibinfo{author}{\bibfnamefont{J.~C.} \bibnamefont{Rojas}}, \bibnamefont{and}
  \bibinfo{author}{\bibfnamefont{Y.}~\bibnamefont{Zhang}},
  \bibinfo{journal}{Nucl. Phys. B Proc. Suppl.}
  \textbf{\bibinfo{volume}{207-208}}, \bibinfo{pages}{273}
  (\bibinfo{year}{2010}), \eprint{1009.1169}.

\bibitem[{\citenamefont{Dominguez
  et~al.}(2013{\natexlab{b}})\citenamefont{Dominguez, Loewe, and
  Zhang}}]{Dominguez:2013fca}
\bibinfo{author}{\bibfnamefont{C.~A.} \bibnamefont{Dominguez}},
  \bibinfo{author}{\bibfnamefont{M.}~\bibnamefont{Loewe}}, \bibnamefont{and}
  \bibinfo{author}{\bibfnamefont{Y.}~\bibnamefont{Zhang}},
  \bibinfo{journal}{Phys. Rev. D} \textbf{\bibinfo{volume}{88}},
  \bibinfo{pages}{054015} (\bibinfo{year}{2013}{\natexlab{b}}),
  \eprint{1307.5766}.

\bibitem[{\citenamefont{Martin~Contreras
  et~al.}(2021)\citenamefont{Martin~Contreras, Diles, and
  Vega}}]{MartinContreras:2021bis}
\bibinfo{author}{\bibfnamefont{M.~A.} \bibnamefont{Martin~Contreras}},
  \bibinfo{author}{\bibfnamefont{S.}~\bibnamefont{Diles}}, \bibnamefont{and}
  \bibinfo{author}{\bibfnamefont{A.}~\bibnamefont{Vega}},
  \bibinfo{journal}{Phys. Rev. D} \textbf{\bibinfo{volume}{103}},
  \bibinfo{pages}{086008} (\bibinfo{year}{2021}), \eprint{2101.06212}.

\bibitem[{\citenamefont{Veli~Veliev et~al.}(2012)\citenamefont{Veli~Veliev,
  Azizi, Sundu, and Kaya}}]{VeliVeliev:2012cc}
\bibinfo{author}{\bibfnamefont{E.}~\bibnamefont{Veli~Veliev}},
  \bibinfo{author}{\bibfnamefont{K.}~\bibnamefont{Azizi}},
  \bibinfo{author}{\bibfnamefont{H.}~\bibnamefont{Sundu}}, \bibnamefont{and}
  \bibinfo{author}{\bibfnamefont{G.}~\bibnamefont{Kaya}}
  (\bibinfo{year}{2012}), \eprint{1205.5703}.

\bibitem[{\citenamefont{Florkowski and Friman}(1994)}]{Florkowski:1993bq}
\bibinfo{author}{\bibfnamefont{W.}~\bibnamefont{Florkowski}} \bibnamefont{and}
  \bibinfo{author}{\bibfnamefont{B.~L.} \bibnamefont{Friman}},
  \bibinfo{journal}{Z. Phys. A} \textbf{\bibinfo{volume}{347}},
  \bibinfo{pages}{271} (\bibinfo{year}{1994}).

\bibitem[{\citenamefont{Bazavov et~al.}(2015)\citenamefont{Bazavov, Karsch,
  Maezawa, Mukherjee, and Petreczky}}]{Bazavov:2014cta}
\bibinfo{author}{\bibfnamefont{A.}~\bibnamefont{Bazavov}},
  \bibinfo{author}{\bibfnamefont{F.}~\bibnamefont{Karsch}},
  \bibinfo{author}{\bibfnamefont{Y.}~\bibnamefont{Maezawa}},
  \bibinfo{author}{\bibfnamefont{S.}~\bibnamefont{Mukherjee}},
  \bibnamefont{and}
  \bibinfo{author}{\bibfnamefont{P.}~\bibnamefont{Petreczky}},
  \bibinfo{journal}{Phys. Rev. D} \textbf{\bibinfo{volume}{91}},
  \bibinfo{pages}{054503} (\bibinfo{year}{2015}), \eprint{1411.3018}.

\bibitem[{\citenamefont{Karsch et~al.}(2012)\citenamefont{Karsch, Laermann,
  Mukherjee, and Petreczky}}]{Karsch:2012na}
\bibinfo{author}{\bibfnamefont{F.}~\bibnamefont{Karsch}},
  \bibinfo{author}{\bibfnamefont{E.}~\bibnamefont{Laermann}},
  \bibinfo{author}{\bibfnamefont{S.}~\bibnamefont{Mukherjee}},
  \bibnamefont{and}
  \bibinfo{author}{\bibfnamefont{P.}~\bibnamefont{Petreczky}},
  \bibinfo{journal}{Phys. Rev. D} \textbf{\bibinfo{volume}{85}},
  \bibinfo{pages}{114501} (\bibinfo{year}{2012}), \eprint{1203.3770}.

\bibitem[{\citenamefont{Gell-Mann}(1962)}]{GellMann:1962xb}
\bibinfo{author}{\bibfnamefont{M.}~\bibnamefont{Gell-Mann}},
  \bibinfo{journal}{Phys. Rev.} \textbf{\bibinfo{volume}{125}},
  \bibinfo{pages}{1067} (\bibinfo{year}{1962}).

\bibitem[{\citenamefont{Ida and Kobayashi}(1966)}]{Ida:1966ev}
\bibinfo{author}{\bibfnamefont{M.}~\bibnamefont{Ida}} \bibnamefont{and}
  \bibinfo{author}{\bibfnamefont{R.}~\bibnamefont{Kobayashi}},
  \bibinfo{journal}{Prog. Theor. Phys.} \textbf{\bibinfo{volume}{36}},
  \bibinfo{pages}{846} (\bibinfo{year}{1966}).

\bibitem[{\citenamefont{Lichtenberg and Tassie}(1967)}]{Lichtenberg:1967zz}
\bibinfo{author}{\bibfnamefont{D.~B.} \bibnamefont{Lichtenberg}}
  \bibnamefont{and} \bibinfo{author}{\bibfnamefont{L.~J.}
  \bibnamefont{Tassie}}, \bibinfo{journal}{Phys. Rev.}
  \textbf{\bibinfo{volume}{155}}, \bibinfo{pages}{1601} (\bibinfo{year}{1967}).

\bibitem[{\citenamefont{Cahill et~al.}(1987)\citenamefont{Cahill, Roberts, and
  Praschifka}}]{Cahill:1987qr}
\bibinfo{author}{\bibfnamefont{R.~T.} \bibnamefont{Cahill}},
  \bibinfo{author}{\bibfnamefont{C.~D.} \bibnamefont{Roberts}},
  \bibnamefont{and}
  \bibinfo{author}{\bibfnamefont{J.}~\bibnamefont{Praschifka}},
  \bibinfo{journal}{Phys. Rev. D} \textbf{\bibinfo{volume}{36}},
  \bibinfo{pages}{2804} (\bibinfo{year}{1987}).

\bibitem[{\citenamefont{Cahill et~al.}(1989)\citenamefont{Cahill, Roberts, and
  Praschifka}}]{Cahill:1988dx}
\bibinfo{author}{\bibfnamefont{R.~T.} \bibnamefont{Cahill}},
  \bibinfo{author}{\bibfnamefont{C.~D.} \bibnamefont{Roberts}},
  \bibnamefont{and}
  \bibinfo{author}{\bibfnamefont{J.}~\bibnamefont{Praschifka}},
  \bibinfo{journal}{Austral. J. Phys.} \textbf{\bibinfo{volume}{42}},
  \bibinfo{pages}{129} (\bibinfo{year}{1989}).

\bibitem[{\citenamefont{Oettel et~al.}(1998)\citenamefont{Oettel, Hellstern,
  Alkofer, and Reinhardt}}]{Oettel:1998bk}
\bibinfo{author}{\bibfnamefont{M.}~\bibnamefont{Oettel}},
  \bibinfo{author}{\bibfnamefont{G.}~\bibnamefont{Hellstern}},
  \bibinfo{author}{\bibfnamefont{R.}~\bibnamefont{Alkofer}}, \bibnamefont{and}
  \bibinfo{author}{\bibfnamefont{H.}~\bibnamefont{Reinhardt}},
  \bibinfo{journal}{Phys. Rev. C} \textbf{\bibinfo{volume}{58}},
  \bibinfo{pages}{2459} (\bibinfo{year}{1998}), \eprint{nucl-th/9805054}.

\bibitem[{\citenamefont{Gutiérrez-Guerrero
  et~al.}(2019)\citenamefont{Gutiérrez-Guerrero, Bashir, Bedolla, and
  Santopinto}}]{Gutierrez-Guerrero:2019uwa}
\bibinfo{author}{\bibfnamefont{L.~X.} \bibnamefont{Gutiérrez-Guerrero}},
  \bibinfo{author}{\bibfnamefont{A.}~\bibnamefont{Bashir}},
  \bibinfo{author}{\bibfnamefont{M.~A.} \bibnamefont{Bedolla}},
  \bibnamefont{and}
  \bibinfo{author}{\bibfnamefont{E.}~\bibnamefont{Santopinto}},
  \bibinfo{journal}{Phys. Rev.} \textbf{\bibinfo{volume}{D100}},
  \bibinfo{pages}{114032} (\bibinfo{year}{2019}), \eprint{1911.09213}.

\bibitem[{\citenamefont{Guti\'errez-Guerrero
  et~al.}(2021)\citenamefont{Guti\'errez-Guerrero, Paredes-Torres, and
  Bashir}}]{Gutierrez-Guerrero:2021rsx}
\bibinfo{author}{\bibfnamefont{L.~X.} \bibnamefont{Guti\'errez-Guerrero}},
  \bibinfo{author}{\bibfnamefont{G.}~\bibnamefont{Paredes-Torres}},
  \bibnamefont{and} \bibinfo{author}{\bibfnamefont{A.}~\bibnamefont{Bashir}},
  \bibinfo{journal}{Phys. Rev. D} \textbf{\bibinfo{volume}{104}},
  \bibinfo{pages}{094013} (\bibinfo{year}{2021}), \eprint{2109.09058}.

\bibitem[{\citenamefont{Chen et~al.}(2012)\citenamefont{Chen, Chang, Roberts,
  Wan, and Wilson}}]{Chen:2012qr}
\bibinfo{author}{\bibfnamefont{C.}~\bibnamefont{Chen}},
  \bibinfo{author}{\bibfnamefont{L.}~\bibnamefont{Chang}},
  \bibinfo{author}{\bibfnamefont{C.~D.} \bibnamefont{Roberts}},
  \bibinfo{author}{\bibfnamefont{S.}~\bibnamefont{Wan}}, \bibnamefont{and}
  \bibinfo{author}{\bibfnamefont{D.~J.} \bibnamefont{Wilson}},
  \bibinfo{journal}{Few Body Syst.} \textbf{\bibinfo{volume}{53}},
  \bibinfo{pages}{293} (\bibinfo{year}{2012}), \eprint{1204.2553}.

\bibitem[{\citenamefont{Barabanov et~al.}(2021)}]{Barabanov:2020jvn}
\bibinfo{author}{\bibfnamefont{M.~Y.} \bibnamefont{Barabanov}}
  \bibnamefont{et~al.}, \bibinfo{journal}{Prog. Part. Nucl. Phys.}
  \textbf{\bibinfo{volume}{116}}, \bibinfo{pages}{103835}
  (\bibinfo{year}{2021}), \eprint{2008.07630}.

\bibitem[{\citenamefont{Wang et~al.}(2013)\citenamefont{Wang, Liu, Chang,
  Roberts, and Schmidt}}]{Wang:2013wk}
\bibinfo{author}{\bibfnamefont{K.-l.} \bibnamefont{Wang}},
  \bibinfo{author}{\bibfnamefont{Y.-x.} \bibnamefont{Liu}},
  \bibinfo{author}{\bibfnamefont{L.}~\bibnamefont{Chang}},
  \bibinfo{author}{\bibfnamefont{C.~D.} \bibnamefont{Roberts}},
  \bibnamefont{and} \bibinfo{author}{\bibfnamefont{S.~M.}
  \bibnamefont{Schmidt}}, \bibinfo{journal}{Phys. Rev. D}
  \textbf{\bibinfo{volume}{87}}, \bibinfo{pages}{074038}
  (\bibinfo{year}{2013}), \eprint{1301.6762}.

\bibitem[{\citenamefont{Bedolla et~al.}(2020)\citenamefont{Bedolla, Ferretti,
  Roberts, and Santopinto}}]{Bedolla:2019zwg}
\bibinfo{author}{\bibfnamefont{M.~A.} \bibnamefont{Bedolla}},
  \bibinfo{author}{\bibfnamefont{J.}~\bibnamefont{Ferretti}},
  \bibinfo{author}{\bibfnamefont{C.~D.} \bibnamefont{Roberts}},
  \bibnamefont{and}
  \bibinfo{author}{\bibfnamefont{E.}~\bibnamefont{Santopinto}},
  \bibinfo{journal}{Eur. Phys. J. C} \textbf{\bibinfo{volume}{80}},
  \bibinfo{pages}{1004} (\bibinfo{year}{2020}), \eprint{1911.00960}.

\bibitem[{\citenamefont{Gutierrez-Guerrero
  et~al.}(2010)\citenamefont{Gutierrez-Guerrero, Bashir, Cloet, and
  Roberts}}]{Gutierrez-Guerrero:2010waf}
\bibinfo{author}{\bibfnamefont{L.~X.} \bibnamefont{Gutierrez-Guerrero}},
  \bibinfo{author}{\bibfnamefont{A.}~\bibnamefont{Bashir}},
  \bibinfo{author}{\bibfnamefont{I.~C.} \bibnamefont{Cloet}}, \bibnamefont{and}
  \bibinfo{author}{\bibfnamefont{C.~D.} \bibnamefont{Roberts}},
  \bibinfo{journal}{Phys. Rev. C} \textbf{\bibinfo{volume}{81}},
  \bibinfo{pages}{065202} (\bibinfo{year}{2010}), \eprint{1002.1968}.

\bibitem[{\citenamefont{Guti\'errez-Guerrero
  et~al.}(2024)\citenamefont{Guti\'errez-Guerrero, Raya, Albino, and
  Hern\'andez-Pinto}}]{Gutierrez-Guerrero:2024him}
\bibinfo{author}{\bibfnamefont{L.~X.} \bibnamefont{Guti\'errez-Guerrero}},
  \bibinfo{author}{\bibfnamefont{A.}~\bibnamefont{Raya}},
  \bibinfo{author}{\bibfnamefont{L.}~\bibnamefont{Albino}}, \bibnamefont{and}
  \bibinfo{author}{\bibfnamefont{R.~J.} \bibnamefont{Hern\'andez-Pinto}},
  \bibinfo{journal}{Phys. Rev. D} \textbf{\bibinfo{volume}{110}},
  \bibinfo{pages}{074015} (\bibinfo{year}{2024}), \eprint{2409.06057}.

\bibitem[{\citenamefont{Ram{\'\i}rez-Garrido
  et~al.}(2025)\citenamefont{Ram{\'\i}rez-Garrido, Hern{\'a}ndez-Pinto,
  Higuera-Angulo, and Guti{\'e}rrez-Guerrero}}]{Ramirez-Garrido:2025rsu}
\bibinfo{author}{\bibfnamefont{M.~A.} \bibnamefont{Ram{\'\i}rez-Garrido}},
  \bibinfo{author}{\bibfnamefont{R.~J.} \bibnamefont{Hern{\'a}ndez-Pinto}},
  \bibinfo{author}{\bibfnamefont{I.~M.} \bibnamefont{Higuera-Angulo}},
  \bibnamefont{and} \bibinfo{author}{\bibfnamefont{L.~X.}
  \bibnamefont{Guti{\'e}rrez-Guerrero}} (\bibinfo{year}{2025}),
  \eprint{2508.01099}.

\bibitem[{\citenamefont{Aguilar et~al.}(2018)\citenamefont{Aguilar, Binosi,
  Figueiredo, and Papavassiliou}}]{Aguilar:2017dco}
\bibinfo{author}{\bibfnamefont{A.~C.} \bibnamefont{Aguilar}},
  \bibinfo{author}{\bibfnamefont{D.}~\bibnamefont{Binosi}},
  \bibinfo{author}{\bibfnamefont{C.~T.} \bibnamefont{Figueiredo}},
  \bibnamefont{and}
  \bibinfo{author}{\bibfnamefont{J.}~\bibnamefont{Papavassiliou}},
  \bibinfo{journal}{Eur. Phys. J.} \textbf{\bibinfo{volume}{C78}},
  \bibinfo{pages}{181} (\bibinfo{year}{2018}), \eprint{1712.06926}.

\bibitem[{\citenamefont{Binosi and Papavassiliou}(2018)}]{Binosi:2017rwj}
\bibinfo{author}{\bibfnamefont{D.}~\bibnamefont{Binosi}} \bibnamefont{and}
  \bibinfo{author}{\bibfnamefont{J.}~\bibnamefont{Papavassiliou}},
  \bibinfo{journal}{Phys. Rev.} \textbf{\bibinfo{volume}{D97}},
  \bibinfo{pages}{054029} (\bibinfo{year}{2018}), \eprint{1709.09964}.

\bibitem[{\citenamefont{Gao et~al.}(2018)\citenamefont{Gao, Qin, Roberts, and
  Rodriguez-Quintero}}]{Gao:2017uox}
\bibinfo{author}{\bibfnamefont{F.}~\bibnamefont{Gao}},
  \bibinfo{author}{\bibfnamefont{S.-X.} \bibnamefont{Qin}},
  \bibinfo{author}{\bibfnamefont{C.~D.} \bibnamefont{Roberts}},
  \bibnamefont{and}
  \bibinfo{author}{\bibfnamefont{J.}~\bibnamefont{Rodriguez-Quintero}},
  \bibinfo{journal}{Phys. Rev.} \textbf{\bibinfo{volume}{D97}},
  \bibinfo{pages}{034010} (\bibinfo{year}{2018}), \eprint{1706.04681}.

\bibitem[{\citenamefont{Raya et~al.}(2018)\citenamefont{Raya, Bedolla,
  Cobos-Mart\'\i{}nez, and Bashir}}]{Raya:2017ggu}
\bibinfo{author}{\bibfnamefont{K.}~\bibnamefont{Raya}},
  \bibinfo{author}{\bibfnamefont{M.~A.} \bibnamefont{Bedolla}},
  \bibinfo{author}{\bibfnamefont{J.~J.} \bibnamefont{Cobos-Mart\'\i{}nez}},
  \bibnamefont{and} \bibinfo{author}{\bibfnamefont{A.}~\bibnamefont{Bashir}},
  \bibinfo{journal}{Few Body Syst.} \textbf{\bibinfo{volume}{59}},
  \bibinfo{pages}{133} (\bibinfo{year}{2018}), \eprint{1711.00383}.

\bibitem[{\citenamefont{Farias et~al.}(2006)\citenamefont{Farias, Dallabona,
  Krein, and Battistel}}]{Farias:2005cr}
\bibinfo{author}{\bibfnamefont{R.~L.~S.} \bibnamefont{Farias}},
  \bibinfo{author}{\bibfnamefont{G.}~\bibnamefont{Dallabona}},
  \bibinfo{author}{\bibfnamefont{G.}~\bibnamefont{Krein}}, \bibnamefont{and}
  \bibinfo{author}{\bibfnamefont{O.~A.} \bibnamefont{Battistel}},
  \bibinfo{journal}{Phys. Rev.} \textbf{\bibinfo{volume}{C73}},
  \bibinfo{pages}{018201} (\bibinfo{year}{2006}), \eprint{hep-ph/0510145}.

\bibitem[{\citenamefont{Farias et~al.}(2008)\citenamefont{Farias, Dallabona,
  Krein, and Battistel}}]{Farias:2006cs}
\bibinfo{author}{\bibfnamefont{R.~L.~S.} \bibnamefont{Farias}},
  \bibinfo{author}{\bibfnamefont{G.}~\bibnamefont{Dallabona}},
  \bibinfo{author}{\bibfnamefont{G.}~\bibnamefont{Krein}}, \bibnamefont{and}
  \bibinfo{author}{\bibfnamefont{O.~A.} \bibnamefont{Battistel}},
  \bibinfo{journal}{Phys. Rev.} \textbf{\bibinfo{volume}{C77}},
  \bibinfo{pages}{065201} (\bibinfo{year}{2008}), \eprint{hep-ph/0604203}.

\bibitem[{\citenamefont{Bedolla et~al.}(2015)\citenamefont{Bedolla,
  Cobos-Mart\'inez, and Bashir}}]{Bedolla:2015mpa}
\bibinfo{author}{\bibfnamefont{M.~A.} \bibnamefont{Bedolla}},
  \bibinfo{author}{\bibfnamefont{J.~J.} \bibnamefont{Cobos-Mart\'inez}},
  \bibnamefont{and} \bibinfo{author}{\bibfnamefont{A.}~\bibnamefont{Bashir}},
  \bibinfo{journal}{Phys. Rev.} \textbf{\bibinfo{volume}{D92}},
  \bibinfo{pages}{054031} (\bibinfo{year}{2015}), \eprint{1601.05639}.

\bibitem[{\citenamefont{Andronic et~al.}(2018)\citenamefont{Andronic,
  Braun-Munzinger, Redlich, and Stachel}}]{Andronic:2017pug}
\bibinfo{author}{\bibfnamefont{A.}~\bibnamefont{Andronic}},
  \bibinfo{author}{\bibfnamefont{P.}~\bibnamefont{Braun-Munzinger}},
  \bibinfo{author}{\bibfnamefont{K.}~\bibnamefont{Redlich}}, \bibnamefont{and}
  \bibinfo{author}{\bibfnamefont{J.}~\bibnamefont{Stachel}},
  \bibinfo{journal}{Nature} \textbf{\bibinfo{volume}{561}},
  \bibinfo{pages}{321} (\bibinfo{year}{2018}), \eprint{1710.09425}.

\bibitem[{\citenamefont{Bazavov et~al.}(2012)}]{Bazavov:2011nk}
\bibinfo{author}{\bibfnamefont{A.}~\bibnamefont{Bazavov}} \bibnamefont{et~al.},
  \bibinfo{journal}{Phys. Rev. D} \textbf{\bibinfo{volume}{85}},
  \bibinfo{pages}{054503} (\bibinfo{year}{2012}), \eprint{1111.1710}.

\bibitem[{\citenamefont{Salpeter and Bethe}(1951)}]{Salpeter:1951sz}
\bibinfo{author}{\bibfnamefont{E.~E.} \bibnamefont{Salpeter}} \bibnamefont{and}
  \bibinfo{author}{\bibfnamefont{H.~A.} \bibnamefont{Bethe}},
  \bibinfo{journal}{Phys. Rev.} \textbf{\bibinfo{volume}{84}},
  \bibinfo{pages}{1232} (\bibinfo{year}{1951}).

\bibitem[{\citenamefont{Bender et~al.}(1996)\citenamefont{Bender, Roberts, and
  Von~Smekal}}]{Bender:1996bb}
\bibinfo{author}{\bibfnamefont{A.}~\bibnamefont{Bender}},
  \bibinfo{author}{\bibfnamefont{C.~D.} \bibnamefont{Roberts}},
  \bibnamefont{and}
  \bibinfo{author}{\bibfnamefont{L.}~\bibnamefont{Von~Smekal}},
  \bibinfo{journal}{Phys. Lett. B} \textbf{\bibinfo{volume}{380}},
  \bibinfo{pages}{7} (\bibinfo{year}{1996}), \eprint{nucl-th/9602012}.

\bibitem[{\citenamefont{Munczek}(1995)}]{Munczek:1994zz}
\bibinfo{author}{\bibfnamefont{H.~J.} \bibnamefont{Munczek}},
  \bibinfo{journal}{Phys. Rev. D} \textbf{\bibinfo{volume}{52}},
  \bibinfo{pages}{4736} (\bibinfo{year}{1995}), \eprint{hep-th/9411239}.

\bibitem[{\citenamefont{Gell-Mann}(1964)}]{GellMann:1964nj}
\bibinfo{author}{\bibfnamefont{M.}~\bibnamefont{Gell-Mann}},
  \bibinfo{journal}{Phys. Lett.} \textbf{\bibinfo{volume}{8}},
  \bibinfo{pages}{214} (\bibinfo{year}{1964}).

\bibitem[{\citenamefont{Zweig}(1964{\natexlab{a}})}]{Zweig:1964jf}
\bibinfo{author}{\bibfnamefont{G.}~\bibnamefont{Zweig}}, in
  \emph{\bibinfo{booktitle}{DEVELOPMENTS IN THE QUARK THEORY OF HADRONS. VOL.
  1. 1964 - 1978}}, edited by
  \bibinfo{editor}{\bibfnamefont{D.}~\bibnamefont{Lichtenberg}}
  \bibnamefont{and} \bibinfo{editor}{\bibfnamefont{S.~P.} \bibnamefont{Rosen}}
  (\bibinfo{year}{1964}{\natexlab{a}}), pp. \bibinfo{pages}{22--101},
  \urlprefix\url{http://inspirehep.net/record/4674/files/cern-th-412.pdf}.

\bibitem[{\citenamefont{Zweig}(1964{\natexlab{b}})}]{Zweig:1964ruk}
\bibinfo{author}{\bibfnamefont{G.}~\bibnamefont{Zweig}}
  (\bibinfo{year}{1964}{\natexlab{b}}).

\bibitem[{\citenamefont{Llewellyn-Smith}(1969)}]{Llewellyn-Smith:1969bcu}
\bibinfo{author}{\bibfnamefont{C.~H.} \bibnamefont{Llewellyn-Smith}},
  \bibinfo{journal}{Annals Phys.} \textbf{\bibinfo{volume}{53}},
  \bibinfo{pages}{521} (\bibinfo{year}{1969}).

\bibitem[{\citenamefont{Maris and Tandy}(1999)}]{Maris:1999nt}
\bibinfo{author}{\bibfnamefont{P.}~\bibnamefont{Maris}} \bibnamefont{and}
  \bibinfo{author}{\bibfnamefont{P.~C.} \bibnamefont{Tandy}},
  \bibinfo{journal}{Phys. Rev. C} \textbf{\bibinfo{volume}{60}},
  \bibinfo{pages}{055214} (\bibinfo{year}{1999}), \eprint{nucl-th/9905056}.

\bibitem[{\citenamefont{Blaschke et~al.}(2001)\citenamefont{Blaschke, Burau,
  Kalinovsky, Maris, and Tandy}}]{Blaschke:2000gd}
\bibinfo{author}{\bibfnamefont{D.}~\bibnamefont{Blaschke}},
  \bibinfo{author}{\bibfnamefont{G.}~\bibnamefont{Burau}},
  \bibinfo{author}{\bibfnamefont{Y.~L.} \bibnamefont{Kalinovsky}},
  \bibinfo{author}{\bibfnamefont{P.}~\bibnamefont{Maris}}, \bibnamefont{and}
  \bibinfo{author}{\bibfnamefont{P.~C.} \bibnamefont{Tandy}},
  \bibinfo{journal}{Int. J. Mod. Phys. A} \textbf{\bibinfo{volume}{16}},
  \bibinfo{pages}{2267} (\bibinfo{year}{2001}), \eprint{nucl-th/0002024}.

\bibitem[{\citenamefont{Navas et~al.}(2024)}]{ParticleDataGroup:2024cfk}
\bibinfo{author}{\bibfnamefont{S.}~\bibnamefont{Navas}} \bibnamefont{et~al.}
  (\bibinfo{collaboration}{Particle Data Group}), \bibinfo{journal}{Phys. Rev.
  D} \textbf{\bibinfo{volume}{110}}, \bibinfo{pages}{030001}
  (\bibinfo{year}{2024}).

\bibitem[{\citenamefont{Montana et~al.}(2022)\citenamefont{Montana, Ramos,
  Tolos, and Torres-Rincon}}]{Montana:2021vks}
\bibinfo{author}{\bibfnamefont{G.}~\bibnamefont{Montana}},
  \bibinfo{author}{\bibfnamefont{A.}~\bibnamefont{Ramos}},
  \bibinfo{author}{\bibfnamefont{L.}~\bibnamefont{Tolos}}, \bibnamefont{and}
  \bibinfo{author}{\bibfnamefont{J.~M.} \bibnamefont{Torres-Rincon}},
  \bibinfo{journal}{EPJ Web Conf.} \textbf{\bibinfo{volume}{259}},
  \bibinfo{pages}{12008} (\bibinfo{year}{2022}), \eprint{2108.04874}.

\bibitem[{\citenamefont{Aarts et~al.}(2022)\citenamefont{Aarts, Allton,
  Bignell, Burns, Garc\'\i{}a-Mascaraque, Hands, J\"ager, Kim, Ryan, and
  Skullerud}}]{Aarts:2022krz}
\bibinfo{author}{\bibfnamefont{G.}~\bibnamefont{Aarts}},
  \bibinfo{author}{\bibfnamefont{C.}~\bibnamefont{Allton}},
  \bibinfo{author}{\bibfnamefont{R.}~\bibnamefont{Bignell}},
  \bibinfo{author}{\bibfnamefont{T.~J.} \bibnamefont{Burns}},
  \bibinfo{author}{\bibfnamefont{S.~C.} \bibnamefont{Garc\'\i{}a-Mascaraque}},
  \bibinfo{author}{\bibfnamefont{S.}~\bibnamefont{Hands}},
  \bibinfo{author}{\bibfnamefont{B.}~\bibnamefont{J\"ager}},
  \bibinfo{author}{\bibfnamefont{S.}~\bibnamefont{Kim}},
  \bibinfo{author}{\bibfnamefont{S.~M.} \bibnamefont{Ryan}}, \bibnamefont{and}
  \bibinfo{author}{\bibfnamefont{J.-I.} \bibnamefont{Skullerud}}
  (\bibinfo{year}{2022}), \eprint{2209.14681}.

\bibitem[{\citenamefont{Song}(1993)}]{Song:1993af}
\bibinfo{author}{\bibfnamefont{C.}~\bibnamefont{Song}}, \bibinfo{journal}{Phys.
  Rev. D} \textbf{\bibinfo{volume}{48}}, \bibinfo{pages}{1375}
  (\bibinfo{year}{1993}).

\bibitem[{\citenamefont{Veliev et~al.}(2011)\citenamefont{Veliev, Azizi, Sundu,
  and Aksit}}]{Veliev:2011zz}
\bibinfo{author}{\bibfnamefont{E.~V.} \bibnamefont{Veliev}},
  \bibinfo{author}{\bibfnamefont{K.}~\bibnamefont{Azizi}},
  \bibinfo{author}{\bibfnamefont{H.}~\bibnamefont{Sundu}}, \bibnamefont{and}
  \bibinfo{author}{\bibfnamefont{N.}~\bibnamefont{Aksit}},
  \bibinfo{journal}{Nucl. Phys. B Proc. Suppl.}
  \textbf{\bibinfo{volume}{219-220}}, \bibinfo{pages}{170}
  (\bibinfo{year}{2011}).

\bibitem[{\citenamefont{Cucchieri et~al.}(2007)\citenamefont{Cucchieri, Maas,
  and Mendes}}]{Cucchieri:2007ta}
\bibinfo{author}{\bibfnamefont{A.}~\bibnamefont{Cucchieri}},
  \bibinfo{author}{\bibfnamefont{A.}~\bibnamefont{Maas}}, \bibnamefont{and}
  \bibinfo{author}{\bibfnamefont{T.}~\bibnamefont{Mendes}},
  \bibinfo{journal}{Phys. Rev. D} \textbf{\bibinfo{volume}{75}},
  \bibinfo{pages}{076003} (\bibinfo{year}{2007}), \eprint{hep-lat/0702022}.

\bibitem[{\citenamefont{Maris and Tandy}(2001)}]{Maris:2001rq}
\bibinfo{author}{\bibfnamefont{P.}~\bibnamefont{Maris}} \bibnamefont{and}
  \bibinfo{author}{\bibfnamefont{P.~C.} \bibnamefont{Tandy}}, in
  \emph{\bibinfo{booktitle}{{Research Program at the Erwin Schodinger Institute
  on Confinement}}} (\bibinfo{year}{2001}), \eprint{nucl-th/0109035}.

\bibitem[{\citenamefont{Bazavov et~al.}(2019)}]{Bazavov:2019www}
\bibinfo{author}{\bibfnamefont{A.}~\bibnamefont{Bazavov}} \bibnamefont{et~al.},
  \bibinfo{journal}{Phys. Rev. D} \textbf{\bibinfo{volume}{100}},
  \bibinfo{pages}{094510} (\bibinfo{year}{2019}), \eprint{1908.09552}.

\bibitem[{\citenamefont{Yazici}(2016)}]{Yazici:2016foi}
\bibinfo{author}{\bibfnamefont{E.}~\bibnamefont{Yazici}}
  (\bibinfo{year}{2016}), \eprint{1605.05289}.

\bibitem[{\citenamefont{Bermudez et~al.}(2017)\citenamefont{Bermudez, Albino,
  Guti\'errez-Guerrero, Tejeda-Yeomans, and Bashir}}]{Bermudez:2017bpx}
\bibinfo{author}{\bibfnamefont{R.}~\bibnamefont{Bermudez}},
  \bibinfo{author}{\bibfnamefont{L.}~\bibnamefont{Albino}},
  \bibinfo{author}{\bibfnamefont{L.~X.} \bibnamefont{Guti\'errez-Guerrero}},
  \bibinfo{author}{\bibfnamefont{M.~E.} \bibnamefont{Tejeda-Yeomans}},
  \bibnamefont{and} \bibinfo{author}{\bibfnamefont{A.}~\bibnamefont{Bashir}},
  \bibinfo{journal}{Phys. Rev. D} \textbf{\bibinfo{volume}{95}},
  \bibinfo{pages}{034041} (\bibinfo{year}{2017}), \eprint{1702.04437}.

\bibitem[{\citenamefont{Bashir et~al.}(2012)\citenamefont{Bashir, Bermudez,
  Chang, and Roberts}}]{Bashir:2011dp}
\bibinfo{author}{\bibfnamefont{A.}~\bibnamefont{Bashir}},
  \bibinfo{author}{\bibfnamefont{R.}~\bibnamefont{Bermudez}},
  \bibinfo{author}{\bibfnamefont{L.}~\bibnamefont{Chang}}, \bibnamefont{and}
  \bibinfo{author}{\bibfnamefont{C.~D.} \bibnamefont{Roberts}},
  \bibinfo{journal}{Phys. Rev. C} \textbf{\bibinfo{volume}{85}},
  \bibinfo{pages}{045205} (\bibinfo{year}{2012}), \eprint{1112.4847}.

\bibitem[{\citenamefont{Chang et~al.}(2011)\citenamefont{Chang, Liu, and
  Roberts}}]{Chang:2010hb}
\bibinfo{author}{\bibfnamefont{L.}~\bibnamefont{Chang}},
  \bibinfo{author}{\bibfnamefont{Y.-X.} \bibnamefont{Liu}}, \bibnamefont{and}
  \bibinfo{author}{\bibfnamefont{C.~D.} \bibnamefont{Roberts}},
  \bibinfo{journal}{Phys. Rev. Lett.} \textbf{\bibinfo{volume}{106}},
  \bibinfo{pages}{072001} (\bibinfo{year}{2011}), \eprint{1009.3458}.

\bibitem[{\citenamefont{Chang and Roberts}(2012)}]{Chang:2011ei}
\bibinfo{author}{\bibfnamefont{L.}~\bibnamefont{Chang}} \bibnamefont{and}
  \bibinfo{author}{\bibfnamefont{C.~D.} \bibnamefont{Roberts}},
  \bibinfo{journal}{Phys. Rev. C} \textbf{\bibinfo{volume}{85}},
  \bibinfo{pages}{052201} (\bibinfo{year}{2012}), \eprint{1104.4821}.

\bibitem[{\citenamefont{Lu et~al.}(2017)\citenamefont{Lu, Chen, Roberts,
  Segovia, Xu, and Zong}}]{Lu:2017cln}
\bibinfo{author}{\bibfnamefont{Y.}~\bibnamefont{Lu}},
  \bibinfo{author}{\bibfnamefont{C.}~\bibnamefont{Chen}},
  \bibinfo{author}{\bibfnamefont{C.~D.} \bibnamefont{Roberts}},
  \bibinfo{author}{\bibfnamefont{J.}~\bibnamefont{Segovia}},
  \bibinfo{author}{\bibfnamefont{S.-S.} \bibnamefont{Xu}}, \bibnamefont{and}
  \bibinfo{author}{\bibfnamefont{H.-S.} \bibnamefont{Zong}},
  \bibinfo{journal}{Phys. Rev.} \textbf{\bibinfo{volume}{C96}},
  \bibinfo{pages}{015208} (\bibinfo{year}{2017}), \eprint{1705.03988}.

\bibitem[{\citenamefont{Yin et~al.}(2021)\citenamefont{Yin, Cui, Roberts, and
  Segovia}}]{Yin:2021uom}
\bibinfo{author}{\bibfnamefont{P.-L.} \bibnamefont{Yin}},
  \bibinfo{author}{\bibfnamefont{Z.-F.} \bibnamefont{Cui}},
  \bibinfo{author}{\bibfnamefont{C.~D.} \bibnamefont{Roberts}},
  \bibnamefont{and} \bibinfo{author}{\bibfnamefont{J.}~\bibnamefont{Segovia}},
  \bibinfo{journal}{Eur. Phys. J. C} \textbf{\bibinfo{volume}{81}},
  \bibinfo{pages}{327} (\bibinfo{year}{2021}), \eprint{2102.12568}.

\bibitem[{\citenamefont{Detar and Kogut}(1987)}]{Detar:1987kae}
\bibinfo{author}{\bibfnamefont{C.~E.} \bibnamefont{Detar}} \bibnamefont{and}
  \bibinfo{author}{\bibfnamefont{J.~B.} \bibnamefont{Kogut}},
  \bibinfo{journal}{Phys. Rev. Lett.} \textbf{\bibinfo{volume}{59}},
  \bibinfo{pages}{399} (\bibinfo{year}{1987}).

\bibitem[{\citenamefont{Wilczek}(2004)}]{Wilczek:2004im}
\bibinfo{author}{\bibfnamefont{F.}~\bibnamefont{Wilczek}}, in
  \emph{\bibinfo{booktitle}{{Deserfest: A Celebration of the Life and Works of
  Stanley Deser}}} (\bibinfo{year}{2004}), pp. \bibinfo{pages}{322--338},
  \eprint{hep-ph/0409168}.

\bibitem[{\citenamefont{Mo et~al.}(2010)\citenamefont{Mo, Qin, and
  Liu}}]{Mo:2010zza}
\bibinfo{author}{\bibfnamefont{Y.}~\bibnamefont{Mo}},
  \bibinfo{author}{\bibfnamefont{S.-x.} \bibnamefont{Qin}}, \bibnamefont{and}
  \bibinfo{author}{\bibfnamefont{Y.-x.} \bibnamefont{Liu}},
  \bibinfo{journal}{Phys. Rev. C} \textbf{\bibinfo{volume}{82}},
  \bibinfo{pages}{025206} (\bibinfo{year}{2010}), \eprint{1009.3314}.

\end{thebibliography}
\end{document}